\documentclass[unpublished, a4paper,onecolumn,11pt]{quantumarticle}
\pdfoutput=1
\usepackage[T1]{fontenc}
\usepackage[english]{babel}
\usepackage[mode=build]{standalone}
\usepackage{graphicx, amsmath, amsthm, braket, mathtools, amsfonts, dsfont, amssymb, tikz, lipsum, quantikz, todonotes, xparse, multirow}
\usepackage[numbers,compress]{natbib}
\usepackage[normalem]{ulem}
\usepackage{hyperref}
\definecolor{adiyellow}{RGB}{200,150,0}    
\definecolor{davidblue}{RGB}{0,0,200}    
\definecolor{petrosgreen}{RGB}{0,200,0} 

\theoremstyle{plain}
\newtheorem{theorem}{Theorem}[section]
\newtheorem{lemma}[theorem]{Lemma}

\newtheorem{corollary}[theorem]{Corollary}

\theoremstyle{definition}
\newtheorem{remark}[theorem]{Remark}

\numberwithin{equation}{section}
\usepackage{thm-restate}
\newcommand{\Z}{\mathbb{Z}}
\newcommand{\qft}[4][]{\mathbf{QFT}_{#2,#3}(#4\if\relax\detokenize{#1}\relax\else, #1\fi)} 
\newcommand{\p}[5][]{\mathbf{P}_{#2, #3, #4}(#5\if\relax\detokenize{#1}\relax\else, #1\fi)}
\newcommand{\pover}[5][]{\overline{\mathbf{P}}_{#2, #3, #4}(#5\if\relax\detokenize{#1}\relax\else, #1\fi)}

\newcommand{\rev}[1]{\overline{#1}}

\allowdisplaybreaks
\title{Fast, in-place, and ancilla-free quantum algorithms for integer multiplication and inversion}
\newcommand{\bylinemark}[1]{\texorpdfstring{\textsuperscript{#1}}{}}
\makeatletter
\AddToHook{cmd/@printtitle/before}{\color{quantumviolet}}
\AddToHook{cmd/@printauthorextrainfo/before}{%
  \let\bylinemark\@gobble
  \hypersetup{urlcolor=quantumviolet}}
\makeatother

\author{David Halpenny\bylinemark{1,*}}
\email{david.halpenny1@outlook.com}

\author{Adithya Sireesh\bylinemark{1,*}}
\email{s1616497@ed.ac.uk}

\author{Petros Wallden\bylinemark{1}}
\affiliation{%
  \textsuperscript{1}Quantum Software Lab, School of Informatics, University of Edinburgh\\
  \textsuperscript{*}These authors contributed equally to this work.\\}
\begin{document}
\maketitle
\begin{abstract}
    Arithmetic operations are a fundamental building block in the design of quantum algorithms, with their implementation cost expected to have a significant impact on the viability of a wide variety of quantum applications. In this work, we present novel algorithms for a number of in-place quantum arithmetic operations that are highly efficient in terms of their space and gate count overheads compared to the current state-of-the-art. We show that in-place modular multiplication of integers, modulo a power of two, can be performed with either a classical or quantum odd multiplier using $O(n^{1+\epsilon})$ gates and no ancilla qubits, for any $\epsilon >0$. We then show that standard in-place integer multiplication can be performed with a non-zero classical or quantum multiplier using $O(n^{1+\epsilon})$ gates for any $\epsilon>0$, with no ancilla qubits required for the classical multiplier and a logarithmic number of ancilla qubits for the quantum. Using our quantum modular multiplication algorithm, we also show that in-place modular inversion, modulo a power of two, can be performed using $O(n^{1+\epsilon})$ gates and no ancilla qubits, for any $\epsilon>0$. To our knowledge, these are the first quantum algorithms achieving both subquadratic gate counts and sublinear ancilla counts for each of the considered in-place operations. All but one of these fast operations are optimal in terms of memory, only requiring enough space to store the output of the operation, which overwrites the input directly. We present our multiplication algorithms as consequences of a single unified theorem, which may find further application within quantum arithmetic.
\end{abstract}
\section{Intro \& Main Results}
Consider an injective function $f:\Z_{2^n} \rightarrow \Z_{2^m}$. A fundamental task across quantum algorithms research is to determine whether various injective functions like $f$ can be computed efficiently using quantum circuits, and if they can, at what resource cost. Gate, depth, and space requirements are the principal resources of interest and are often traded for one another. For well-studied functions, such as certain arithmetic operations, asymptotic performance improvements are highly sought after and heavily researched, due to the widespread use of these operations as subroutines in larger algorithms \cite{Bacon_2006, hallgren2007, Kassal_2008, Babbush_2019, regev2024efficientquantumfactoringalgorithm, Rubin_2024, Shor_1997, proos2004shorsdiscretelogarithmquantum, Jordan_2025}.

Two natural approaches for computing these functions exist, typically referred to as \emph{in-place} and \emph{out-of-place}, illustrated (without considering potential ancillas) below:
\begin{equation}
\begin{aligned}
    \text{out-of-place}:     & \ \ket{x}_n\ket{0}_m \xmapsto{\hspace{2.25em}} \ket{x}_n\ket{f(x)}_m \\
    \text{in-place}: &\ \ket{x}_n\ket{0}_{m-n} \xmapsto{\hspace{1.2em}} \ket{f(x)}_m.
\end{aligned}
\end{equation}
In the context of a larger algorithm, each approach offers a distinct benefit. If the function input needs to be reused later, retaining it with an out-of-place approach is desirable. Otherwise, an in-place approach, defined by its \emph{uncomputation} of the function input is preferable. In many algorithms, uncomputation is an essential component~\cite{Shor_1997, proos2004shorsdiscretelogarithmquantum, Jordan_2025, regev2009lattices, Berry:2009cwy, Ambainis_2011, berry2019qubitization}.

The standard approach for out-of-place integer multiplication in both classical and quantum settings is the schoolbook method, which can be implemented by repeatedly using an addition circuit, requiring a gate count that scales quadratically with respect to the length of the input.
\emph{Fast} constructions achieving subquadratic gate counts have also been developed, based on recursive multiplication algorithms from classical computing \cite{zalkaFastVersionsShors1998, kowada2006reversible, parentImprovedReversibleQuantum2017, duttaQuantumCircuitsToomCook2018, gidneyAsymptoticallyEfficientQuantum2019, nieQuantumCircuitDesign2023, kahanamoku-meyerFastQuantumInteger2024}. Until recently, it was a major open question whether the recursive nature of these constructions could be reconciled with quantum reversibility without incurring a space overhead, for the storage of intermediate computations, scaling linearly in the length of the input. This question was resolved in \cite{kahanamoku-meyerFastQuantumInteger2024}, when it was proven that out-of-place quantum integer multiplication is possible using $O(n^{1+\epsilon})$ gates and no ancilla qubits, for any $\epsilon>0$.

It is a well-known fact that two unitaries $U_{f}$ and $U_g$ computing $f$ and its left-inverse $g$ out-of-place can be used to build an in-place algorithm for $f$ \cite{Bennett1973LogicalReversibility}. This is clear by considering the following sequence of operations:
\begin{equation}
    \ket{x}_{n}\ket{0}_m \xmapsto{U_{f}} \ket{x}_n\ket{f(x)}_m \xmapsto{\text{SWAP}} \ket{f(x)}_m\ket{x}_n \xmapsto{U_{g}^{-1}} \ket{f(x)}_m\ket{0}_n.
\end{equation}
However, the space efficiency of this in-place algorithm is clearly sub-optimal, involving $n$ ancilla qubits for the simultaneous storage of both the function input and output, in addition to any workspace required. In contrast, an \emph{ancilla-free}, in-place algorithm for computing $f$ is optimally space-efficient, directly overwriting the input with the function output. In-place and ancilla-free constructions for standard multiplication and modular multiplication modulo a power of two exist \cite{gidney2017simple, rines2018high, mogensen2018garbage, seidelEfficientFloatingPoint2021, li2022circuit, fan2024designs} (\cite{mogensen2018garbage} is nearly ancilla-free), all scaling quadratically with respect to the length of the input in terms of gate count.

The history of quantum instantiations for fast multiplication and the evident tension between gate and space cost motivate the following question: since in-place and ancilla-free algorithms reduce space cost maximally, might they be fundamentally prevented from being made fast?

\textbf{Our Contributions.} In this paper, we show that this is not the case. We present in-place and ancilla-free (in all but one of the cases we highlight) algorithms achieving subquadratic gate count for a number of quantum arithmetic operations (see Table \ref{tab:arithmetic-summary-option-4}). 

In more detail, we show that integer multiplication modulo a power of two can be performed in-place on a quantum register between two $O(n)$-bit integers using $O(n^{1+\epsilon})$ gates and no ancilla qubits, for any $\epsilon >0$. The multiplier can be either a classically chosen value $a \in \Z_{2^{n}}^{\times}$ (Theorem \ref{thm:classical mod mult}) or a quantum state $\ket{x}_m$ for $x \in \Z_{2^{m}}^{\times}$ and $m = O(n)$ (Theorem \ref{thm:quantum mod mult}).

We then show that standard integer multiplication can be performed in-place on a quantum register between two $O(n)$-bit integers using $O(n^{1+\epsilon})$ gates for any $\epsilon >0$, and with a multiplier-dependent ancilla count (Theorems \ref{thm:classical full product} and \ref{thm:quantum full product}). When the multiplier is a classically chosen value $a \in \Z_{2^m}\setminus \{0\}$ or a quantum state $\ket{x}_m$ with $x \in \Z_{2^m}^{\times}$ and $m=O(n)$, no ancillas are required, and when $x \in \Z_{2^m}\setminus \{0\}$, just $\lceil \log_2 m \rceil$ ancillas are required. Note that in-place multiplication by zero is not a unitary operation (see Remark \ref{rem:zero-input-case} for details on handling this case). We also prove that our multiplication algorithms involving a classical value have $O(n\log n)$ circuit depth.

Finally, we show that modular inversion of $x \in \Z_{2^n}^{\times}$ can be performed in-place on a quantum register using $O(n^{1+\epsilon})$ gates and $s$ ancilla qubits, for any $\epsilon>0$ and $0\leq s \leq n-1$ (Theorem \ref{thm:budget}). The ancilla count $s$ is a parameter affecting the true gate count of the algorithm and does not impact the stated asymptotic upper bound, meaning that our fast modular inversion algorithm can be ancilla-free.  

\begin{table}[t]
\centering
\begingroup
\vspace{1em}
\small
\setlength{\tabcolsep}{3pt}
\renewcommand{\arraystretch}{1.35}
\begin{tabular*}{\textwidth}{@{\extracolsep{\fill}}r@{\hspace{0.3em}}c@{\hspace{0.3em}}l||c|c|c|c|c@{}}
\hline
\multicolumn{3}{@{}c||}{\rule{0pt}{2.8ex}\textbf{Arithmetic Operation}}
& \textbf{Domain} & \textbf{Gates $G$} & \textbf{Ancillas}
& \textbf{Depth $D$} & \textbf{Theorem}\\[0.5ex]
\hline
$\ket{z}_n$ & $\longmapsto$ & $\ket{az\bmod 2^n}_n$ & $a\in\mathbb{Z}_{2^n}^{\times}$ & $O(n^{1+\epsilon})$ & $\mathbf{0}$ & $O(n\log n)$ & Thm~\ref{thm:classical mod mult}\\
\hline
$\ket{0}_m\ket{z}_n$ & $\longmapsto$ & $\ket{az}_{m+n}$ & $a\in\mathbb{Z}_{2^m}\setminus\{0\}$ & $O(n^{1+\epsilon})$ & $\mathbf{0}$ & $O(n\log n)$ & Thm~\ref{thm:classical full product}\\
\hline
$\ket{x}_m\ket{z}_n$ & $\longmapsto$ & $\ket{x}_m\ket{xz\bmod 2^n}_n$ & $x\in\mathbb{Z}_{2^m}^{\times}$ & $O(n^{1+\epsilon})$ & $\mathbf{0}$ & $\le G$ & Thm~\ref{thm:quantum mod mult}\\
\hline
$\ket{x}_m\ket{0}_m\ket{z}_n$ & $\longmapsto$ & $\ket{x}_m\ket{xz}_{m+n}$ & \begin{tabular}[c]{@{}c@{}}$x\in\mathbb{Z}_{2^m}^{\times}$\\$x\in\mathbb{Z}_{2^m}\setminus\{0\}$\end{tabular} & \begin{tabular}[c]{@{}c@{}}$O(n^{1+\epsilon})$\\$O(n^{1+\epsilon})$\end{tabular} & \begin{tabular}[c]{@{}c@{}}$\mathbf{0}$\\$\lceil\log_2 m\rceil$\end{tabular} & \begin{tabular}[c]{@{}c@{}}$\le G$\\$\le G$\end{tabular} & \begin{tabular}[c]{@{}c@{}}Thm~\ref{thm:quantum full product}\\Thm~\ref{thm:quantum full product}\end{tabular}\\
\hline
$\ket{x}_n$ & $\longmapsto$ & $\ket{x^{-1}\bmod 2^n}_n$ & $x\in\mathbb{Z}_{2^n}^{\times}$ & $O(n^{1+\epsilon})$ & $s$ & $\le G$ & Thm~\ref{thm:budget}\\
\hline
\end{tabular*}
\caption{\textbf{Main arithmetic results.} The results hold for any $\epsilon> 0$ as long as $m=O(n)$. The first two results are for multiplication (modular and standard) of a quantum register by a classical multiplier. The next two are for multiplication (modular and standard) between two quantum registers. The fifth result is for modular inversion of a quantum register. The ancilla count $s$ is a parameter in our construction that determines the exact cost of the algorithm. For all $s$ in the range $0\le s\le n-1$ the claimed asymptotic gate count holds. We work in the abstract quantum circuit model.}
\label{tab:arithmetic-summary-option-4}
\endgroup
\end{table}

To the best of our knowledge, these are the first known quantum algorithms achieving both subquadratic gate counts and sublinear ancilla counts for all of the stated in-place operations. With the exception of our algorithm for standard multiplication by $\ket{x}_m$ for $x \in \Z_{2^m}\setminus{\{0\}}$, all of these results are optimal in terms of space cost.

We unify our multiplication algorithms using a single construction (Theorem \ref{thm:main theorem}) that treats quantum versus classical and modular versus standard multiplication as parameters that can be turned on and off before running the algorithm. The results that we highlight in Table \ref{tab:arithmetic-summary-option-4} are therefore a selection of well-motivated operations in the context of quantum arithmetic following from this main theorem, and not an exhaustive list. We believe this main theorem of our work may help to build other fast and space-efficient arithmetic operations.

Central to our results are two subroutines: the phase-encoded fast integer multiplication operation introduced in \cite{kahanamoku-meyerFastQuantumInteger2024}, which we extend (Lemmas \ref{lem:new P gate complexity two inputs} and \ref{lem:new P gate complexity three inputs}), and a fast controlled and parametrised version of the quantum Fourier transform, which to our knowledge we introduce here for the first time, and which we expect to be of independent interest (Lemma \ref{lem:qft gate complexity}).

Motivated by highly space-constrained settings where temporarily idle qubits may be repurposed \cite{berry2019qubitization, gidney2017factoringdirty, haner2016factoring, buhrman2025quantum},
we develop variants of our standard integer multiplication algorithms which encode the $(m+n)$-bit product on an idle quantum register in a potentially unknown state (Theorems \ref{thm: classical dirty product} and \ref{thm:quantum dirty product}). If used within a larger algorithm, these constructions may be even more space-efficient than the main results of this work, as they avoid initialising an $m$ qubit register $\ket{0}_m$ to store the $(m+n)$-bit product (see Remark \ref{rem:dirty qubits motivation} for more details). 

\subsection{Paper Outline}
Our work is structured into four main sections. In Section \ref{sec:main theorem}, we introduce the fast and ancilla-free subroutines critical for this work and we present our main theorem that unifies standard and modular multiplication with classical and quantum multipliers in a single construction. We prove the cost of these subroutines in Appendices \ref{sec:appendix:Pgate} and \ref{sec:appendix:qft}, and an extension of our main theorem in Appendix \ref{sec:appendix:extended main theorem}. In Section \ref{sec:classical multiplication}, we present our fast, in-place, and ancilla-free algorithms for integer multiplication with a classical multiplier, and in Section \ref{sec:quantum-int-mul}, we present our algorithms for multiplication with a quantum multiplier. In Section \ref{sec:inversion}, we use our algorithm for modular multiplication, modulo a power of two, to construct a fast algorithm for inversion modulo a power of two. We leave the proof of some additional subroutines used in our multiplication and inversion algorithms for Appendices \ref{sec:appendix:additional} and \ref{sec:appendix:inversion}. We conclude with a discussion in Section \ref{sec:discussion}.

\subsection{Conventions}
Throughout this work we adopt the following conventions. A length $n$ bitstring is denoted $x_{n-1}\cdots x_0$ and identified with the integer $x = \sum_{i=0}^{n-1} 2^{i}x_i$ in $\Z_{2^n}$. The bit-reversal $x_{0} \cdots x_{n-1}$ is denoted $\rev{x}$ and identified with $\sum_{i=0}^{n-1}2^{n-i-1}x_i$. The variables $w,x,y,z$ will always refer to integers that are stored within a quantum register. We refer to these as quantum integers. The variables $a,b, c$ will always refer to an integer that is a classical parameter of an algorithm, chosen before the algorithm starts. We refer to this as a classical integer. We use $\Z_{2^n}^\times$ to refer to the multiplicative group of $\Z_{2^n}$, whose elements we identify with odd integers.

We use subscripts to denote the number of qubits contained within a quantum register and to denote the number of qubits a unitary operation is being applied to. For example, $\ket{x}_n$ refers to an $n$ qubit quantum register storing the integer $x$ and $U_{m,n}$ refers to a unitary transformation applied to two registers of sizes $m$ and $n$. We use $\mathbf{I}_{m}$ for the identity on $m$ qubits. If during an algorithm, we apply an operation to just a subregister of the total register of qubits, it is understood that the rest of the qubits are receiving the identity. 

Unless stated otherwise logarithms are taken base 2.
We define the function $\mathrm{tz}: \Z_{2^n} \rightarrow \Z_{n+1}$ by $\mathrm{tz}(x) = \log(\gcd(x, 2^n))$. This function counts the number of trailing zeroes of the bitstring $x$, evaluating to $n$ for the zero bitstring.
We use $G(\cdot)$ and $D(\cdot)$ to denote the gate count and depth of an operation. When solving gate and depth recurrence relations, we often implicitly apply the master theorem for divide and conquer algorithms.
We use $\mathds{1}(P)$ to denote an indicator function which takes the value 1 when the predicate $P$ is true and 0 when it is not.

We work in the abstract quantum circuit model. We assign equal cost to all elementary gates and assume that arbitrary rotation gates can be implemented exactly. We assume qubits have all-to-all connectivity and that gates acting on disjoint sets of qubits can be executed in parallel.

\section{Main Theorem}\label{sec:main theorem}
We now develop the subroutines used in our main theorem. We present algorithms for phase-encoded integer multiplication between two or three differently sized registers and establish their gate, depth, and ancilla costs. We use these operations to develop a fast controlled and parametrised quantum Fourier transform that requires no ancilla qubits. Together, these results provide the tools needed to treat multiplication by both classical and quantum integers within a single construction, culminating in our main theorem.
\subsection{Subroutines}
We start by defining the operations that we need for our main theorem, Theorem \ref{thm:main theorem}. First, we consider the parametrised operations $\p[\alpha]{t}{m}{n}{\phi}, \pover[\alpha]{t}{m}{n}{\phi}$ acting on $t+m+n$ qubits defined as follows. 
Let $x \in \Z_{2^t}, y \in \Z_{2^{m}}, z \in \Z_{2^n}, \phi \in \mathbb{Q}$ and $\alpha \in \{0,1\}$. Then
\begin{align}
\p[\alpha]{t}{m}{n}{\phi}:\ket{x}_t\ket{y}_m\ket{z}_{n} &\longmapsto e^{2\pi i \phi yzx^{\alpha}} \ket{x}_t\ket{y}_m\ket{z}_n, \\ \pover[\alpha]{t}{m}{n}{\phi}: \ket{x}_t\ket{y}_m\ket{z}_{n} &\longmapsto e^{2\pi i \phi\rev{y}zx^{\alpha}} \ket{x}_t\ket{y}_m\ket{z}_{n}.
\end{align}
Both gates perform a multiplication of integers in the phase. When $\alpha=1$ there are three integers involved and when $\alpha=0$ there are just two. We use the subscripts $t,m$ and $n$ to indicate the sizes of the registers that we are applying the gate to. The first of the subscripts indicates which register is optional, i.e. controlled by $\alpha$. The parameter $\phi$ determines the fraction that appears in the phase of the complex exponential. The gate is well-defined as long as $t,m,n \geq 0$ and at least one of these is non-zero. Since it serves no purpose, we drop the parameter $\alpha$ from the definition if $t=0$. It is worth observing that $\mathbf{I}_{t} \otimes \p{0}{m}{n}{\phi} = \p[0]{t}{m}{n}{\phi}$.

It is proven in \cite{kahanamoku-meyerFastQuantumInteger2024} that, for any $\epsilon>0$, the operations $\p{0}{n}{n}{\phi}$ and $\p[1]{n}{n}{n}{\phi}$ can be performed using just $O(n^{1+\epsilon})$ gates without requiring any ancilla qubits. These results are reviewed in Appendix \ref{sec:appendix:Pgate}. It is also shown in \cite{kahanamoku-meyerFastQuantumInteger2024} how to improve the circuit depth of these operations using ancillas. The authors observed that, without the use of ancillas, the algorithm for $\p{0}{n}{n}{\phi}$ seems to require a circuit of depth $\Omega(n)$. 

In the following lemma, we show that $\p{0}{n}{n}{\phi}$ can be performed in $O(n)$ depth without the use of ancillas, matching this suggested lower bound.

\begin{restatable}[Linear-depth phase multiplication]{lemma}{newPgatedepth}\label{lem:newPgatedepth}
    Let $n>0$. For any $\epsilon>0$, there exists a quantum circuit for $\p{0}{n}{n}{\phi}$ that uses $O(n^{1+\epsilon})$ gates, $O(n)$ depth, and no ancilla qubits.
\end{restatable}
\begin{proof}[Proof sketch]
    The full proof of this lemma is left for Appendix \ref{sec:appendix:Pgate}. For those familiar with \cite{kahanamoku-meyerFastQuantumInteger2024}, the key observation is that we can choose the $(2k-1)\times k$ matrix $A$ so that, when we divide the input registers $\ket{y}_n$ and $\ket{z}_n$ into $k$ subregisters, at least two linear combinations of these $k$ subregisters can always be prepared simultaneously, using a sequence of out-of-place quantum additions. This roughly halves the number of sequential subproblems at each level of recursion in the algorithm for $\p{0}{n}{n}{\phi}$. As a result, the $O(n)$ depth requirement at the top level of recursion dominates the total algorithm depth. The gate and ancilla count costs are retained. The matrix $A$ that we choose uses evaluation points $\{\infty, 0, 1, 2, \ldots, 2k-3\}$.
\end{proof}

The following two lemmas show how to handle phase-encoded multiplication between registers of different sizes, extending the results of \cite{kahanamoku-meyerFastQuantumInteger2024}. We show that the multiplication costs no more than in the equal-size case, with the register size set to the bit length of the largest input.

\begin{restatable}[Phase multiplication with two differently sized inputs]{lemma}{Pgatecomplexitytwoinputs}\label{lem:new P gate complexity two inputs}
    Let $n>0$ and $0\leq m\leq n$. For any $\epsilon>0$, there exists a quantum circuit for $\p{0}{m}{n}{\phi}$ that uses $O(n^{1+\epsilon})$ gates, $O(n)$ depth, and no ancilla qubits.
\end{restatable}
\begin{proof}[Proof sketch]
    The full proof of this lemma is left for Appendix \ref{sec:appendix:Pgate}. We give a brief sketch of the proof here. Consider $q>0$ and $0\leq r<m$ such that $n=qm+r$. Dividing $\ket{z}_n$ into $q$ subregisters $Z_{q-1}, \ldots, Z_{0}$ of size $m$, and 1 subregister of size $r$, we apply $\mathbf{P}_{0,m,m}(2^{mi}\phi)$ to $\ket{y}_m$ and $Z_i$. This reduces the problem to performing $\p{0}{m}{r}{2^{qm}\phi}$. This construction can be applied recursively, with the register sizes determined by the Euclidean algorithm for finding the greatest common divisor of $n$ and $m$. The claimed asymptotic costs follow.
\end{proof}

\begin{restatable}[Phase multiplication with up to three differently sized inputs]{lemma}{Pgatecomplexitythreeinputs}\label{lem:new P gate complexity three inputs}
    Let $n>0$ and $0\leq t\leq m\leq n$. For any $\epsilon>0$, there exist quantum circuits for $\p[\alpha]{t}{m}{n}{\phi}$ and $\pover[\alpha]{t}{m}{n}{\phi}$ that use $O(n^{1+\epsilon})$ gates and no ancilla qubits. For the special case $t=0$, these circuits have $O(n)$ depth.
\end{restatable}
\begin{proof}[Proof sketch]
    The full proof of this lemma is left for Appendix \ref{sec:appendix:Pgate}. For those familiar with \cite{kahanamoku-meyerFastQuantumInteger2024}, or for those who have read the review we provide in Appendix \ref{sec:appendix:Pgate}, we give a brief sketch of the proof here. The key is to apply three recursions. The first recursion creates $k$ subproblems using the decomposition of $\phi z$ from Equations \ref{eq:P gate decomposition of phi} and is applied to depth $s=\lfloor \log_k (n/m)\rfloor$, until the subproblems are between registers of size $n/k^s, m$ and $t$. The second recursion creates $2k-1$ subproblems using the decomposition of $\phi yz$ from Equations \ref{eq:P gate decomposition of phi} and is applied to depth $r=\lfloor \log_k(n/tk^s)\rfloor$, until the subproblems are between registers of size $n/k^{s+r}, n/k^{s+r}$ and $t$. The final recursion creates $3k-2$ subproblems using the decomposition of $\phi xyz$ and is applied until the subproblems are between registers of size $O(1)$. A key observation is that at the boundary of the first recursion and at the boundary of the second recursion, the register of size $m$ and the register of size $t$, is treated as though it has been padded with zeroes. This virtual padding ensures that we can view the integers that we recurse on next as if they are sitting on registers of equal-size, making the decompositions of Equations \ref{eq:P gate decomposition of phi} applicable. The gate count can be upper bounded by $G(\p[1]{n}{n}{n}{\phi})$ and the algorithms require no ancillas.
\end{proof}

Next, we consider the operation $\qft[\alpha]{m}{n}{\phi}$ acting on $m+n$ qubits defined as follows. Let $x \in \Z_{2^m}^{\times}, y \in \Z_{2^n}$ and $\phi\in \Z_{2^n}^{\times}$ and let $\alpha \in \{0,1\}$ Then
\begin{align}
\qft[\alpha]{m}{n}{\phi}: \ket{x}_m\ket{y}_n \longmapsto \ket{x}_m\frac{1}{\sqrt {2^{n}}} \sum_{z \in \Z_{2^n}}e^{2\pi i \phi yzx^{\alpha}/2^n}\ket{\rev{z}}_n.
\end{align}
This operation is a controlled and parametrised version of the standard quantum Fourier transform on $\Z_{2^n}$ which leaves bitstrings in the output superposition in reverse order. When $\alpha = 1$ the parametrised QFT is controlled on the register of $m$ qubits and when $\alpha =0$ it is not. As before, we use the subscripts $m$ and $n$ to indicate the size of the registers that we are applying the gate to. The first subscript always refers to the optional control register. The gate is well-defined as long as $m\geq 0$ and $n>0$. For $m=0$, we can drop the parameter $\alpha$, as it serves no purpose. It is worth observing that $\qft{0}{n}{1}$ is the standard QFT on $\Z_{2^n}$ and that $\qft[0]{m}{n}{\phi}= \mathbf{I}_{m} \otimes \qft{0}{n}{\phi}$. We note that the parametrised QFT on $\Z_{2^n}$, $\qft{0}{n}{\phi}$ in our notation, was also considered in the context of quantum arithmetic in \cite{rines2018high,seidelEfficientFloatingPoint2021}.
\begin{restatable}[Fast controlled and parametrised QFT]{lemma}{qftgatecomplexity}\label{lem:qft gate complexity}
    Let $n>0$ and $m =O(n)$. For any $\epsilon >0$, there exists a quantum circuit for $\qft[\alpha]{m}{n}{\phi}$ that uses $O(n^{1+\epsilon})$ gates and no ancilla qubits. For the special case $m=0$, the circuit has $O(n\log n)$ depth.
\end{restatable}
\begin{proof}[Proof sketch]
    We outline the proof of this lemma, which is left for Appendix \ref{sec:appendix:qft}. We give a recursive definition for $\qft{0}{n}{\phi}$, resembling the recursive definition for the standard QFT on $\Z_{2^n}$ found in \cite{cleveFastParallelCircuits2000}. Using Lemma \ref{lem:new P gate complexity two inputs}, we prove the claim for $m=0$. We then consider the control register and give a similar recursive definition for $\qft[1]{m}{n}{\phi}$. Using this definition and Lemma \ref{lem:new P gate complexity three inputs}, we prove the claim for $m=O(n)$. 
\end{proof}
Our discovery that a controlled and parametrised version of the QFT can be made fast, without requiring any ancilla qubits, is fundamental, directly enabling in-place multiplication algorithms that are fast and ancilla-free. We note that, in \cite{kahanamoku-meyerFastQuantumInteger2024}, the standard QFT on $\Z_{2^n}$, $\qft{0}{n}{1}$ in our notation, was shown to require $O(n^{1+\epsilon})$ gates and no ancilla qubits, for any $\epsilon>0$, using the recursive definition of the QFT from \cite{cleveFastParallelCircuits2000}. 

Now that we have introduced these two operations, $\p[\alpha]{t}{m}{n}{\phi}$ and $\qft[\alpha]{m}{n}{\phi}$, which are of critical importance to the algorithms in this paper, we proceed to the main theorem of this work, from which many of our results will follow.

\subsection{Main Theorem}

\begin{theorem}[Main Theorem]\label{thm:main theorem}
    Let $m>0$, $s = O(m)$ and $t=O(m)$. Let $w \in \Z_{2^s}^{\times}$, $x \in \Z_{2^t}$, $y \in \Z_{2^{m}}$, $b, c \in \Z \setminus 2\Z$, and $\alpha, \beta \in \{0,1\}$. For any $\epsilon >0$, there exists a quantum circuit for performing the operation \begin{equation}
        \ket{w}_{s}\ket{x}_{t}\ket{y}_{m} \longmapsto \ket{w}_{s}\ket{2^{m}bw^{\alpha}x+cw^{\beta}y\bmod 2^{t+m}}_{t+m}
    \end{equation} that uses $O(m^{1 +\epsilon})$ gates and no ancilla qubits. For the special case $s=0$, the circuit has $O(m\log m)$ depth. In this case, we adopt the convention $w^{\alpha} = w^{\beta} =1$.
\end{theorem}
\begin{proof}
We provide an algorithm that proves this claim. On input $\ket{w}_s\ket{x}_t\ket{y}_m$ apply the following operations.
    \begin{flalign}
    &\xrightarrow{\qft[\alpha]{s}{t}{b} \otimes \mathbf{I}_{m}} \ket{w}_s\frac{1}{\sqrt {2^{t}}} \sum_{v \in \Z_{2^{t}}}e^{2\pi i bxvw^{\alpha}/2^{t}}\ket{\rev{v}}_{t}\ket{y}_{m}, &&
    \end{flalign}
    \begin{flalign}
    &\xrightarrow{\pover[\beta]{s}{t}{m}{c/2^{t+m}}} \ket{w}_{s}\frac{1}{\sqrt {2^{t}}} \sum_{v \in \Z_{2^{t}}}e^{2\pi i (bxvw^{\alpha}/2^{t} + cvyw^{\beta}/2^{t+m})}\ket{\rev{v}}_{t}\ket{y}_{m}, &&
    \end{flalign}
    \begin{flalign}
    &\label{eq:quantum mult main pre-qft equation}\xrightarrow{\mathbf{I}_{t} \otimes \qft[\beta]{s}{m}{c}} \ket{w}_{s}\frac{1}{\sqrt {2^{t+m}}} \sum_{v \in \Z_{2^{t}}}e^{2\pi i (bxvw^{\alpha}/2^{t} + cvyw^{\beta}/2^{t+m})}\ket{\rev{v}}_{t}\sum_{z \in \Z_{2^m}}e^{2\pi i (cyzw^{\beta}/2^m)}\ket{\rev{z}}_{m}. &&
    \end{flalign}
    Note that \(\ket{\rev{v}}_{t}\ket{\rev{z}}_{m} = \ket{\rev{2^{t}z + v}}_{t+m}\) and 
    \begin{align}
    \nonumber e^{2\pi i (bxvw^{\alpha}/2^{t} + cvyw^{\beta}/2^{t+m}+ cyzw^{\beta}/2^m)}
    &= e^{2\pi i (2^mbxvw^{\alpha} + 2^{t+m}bxzw^{\alpha}+ cvyw^{\beta}+ 2^tcyzw^{\beta})/2^{t+m}}\\
    &= e^{2\pi i (2^tz + v)(2^{m}bxw^{\alpha} + cyw^{\beta})/2^{t+m}}.
    \end{align} 
    Therefore, we can rewrite Equation \ref{eq:quantum mult main pre-qft equation} as
    \begin{flalign}
    &=\ket{w}_s\frac{1}{\sqrt {2^{t+m}}} \sum_{v \in \Z_{2^{t}}}\sum_{z \in \Z_{2^m}}e^{2\pi i (2^tz + v)(2^{m}bxw^{\alpha} + cyw^{\beta})/2^{t+m}} \ket{\rev{2^{t}z + v}}_{t+m}. &&
    \end{flalign}
    Since $(z,v) \mapsto 2^{t}z + v$ describes a one-to-one correspondence between $\Z_{2^m} \times \Z_{2^{t}}$ and $\Z_{2^{t+m}}$, this equals
    \begin{flalign}
    &=\ket{w}_s\frac{1}{\sqrt {2^{t+m}}} \sum_{q \in \Z_{2^{t+m}}}e^{2\pi i q(2^{m}bxw^{\alpha} + cyw^{\beta})/2^{t+m}} \ket{\rev{q}}_{t+m}. &&
    \end{flalign}
    Then finally apply,
    \begin{flalign}
    &\xrightarrow{\mathbf{I}_{s} \otimes \qft{0}{t+m}{1}^{-1}} \ket{w}_s\ket{2^{m}bw^{\alpha}x + cw^{\beta}y\bmod 2^{t+m}}_{t+m}.&&
    \end{flalign}
    By Lemmas \ref{lem:new P gate complexity three inputs} and \ref{lem:qft gate complexity}, this algorithm can be performed using $O(m^{1+\epsilon})$ gates and no ancilla qubits, for any $\epsilon>0$. For the special case $s=0$, by Lemmas \ref{lem:new P gate complexity two inputs} and \ref{lem:qft gate complexity}, the circuit has depth $O(m\log m)$.
\end{proof}
\begin{figure}[!htp]
    \centering
    \includegraphics[width=\textwidth]{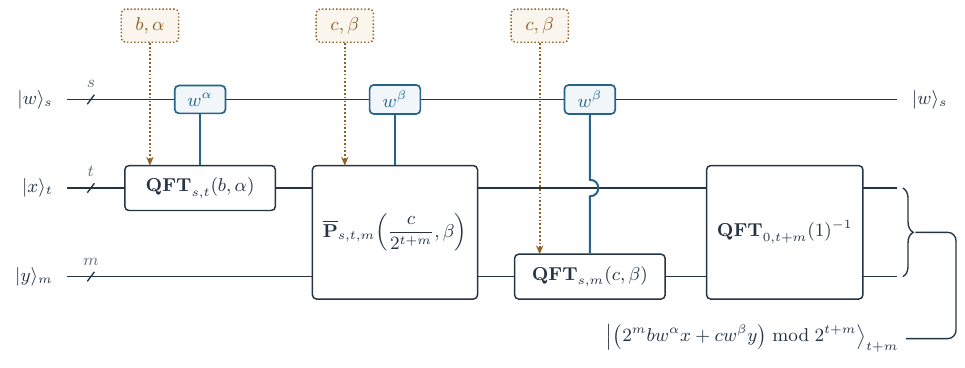}
    \caption{Illustration of the algorithm in Theorem~\ref{thm:main theorem}. Dashed brown boxes refer to classical input parameters. Solid blue boxes refer to optional quantum control.}
    \label{fig:main-theorem-circuit}
\end{figure}
\begin{remark}
In Theorem \ref{thm:main theorem extended} of Appendix~\ref{sec:appendix:extended main theorem}, we extend the domain of the operation in Theorem~\ref{thm:main theorem} to arbitrary quantum multipliers $w\in\mathbb{Z}_{2^s}$. This extension turns out to be useful in the proof of Lemma~\ref{lem:double} in Appendix~\ref{sec:appendix:inversion} and may also be of independent interest.
\end{remark}

We show in the next section that Theorem \ref{thm:main theorem} immediately implies the existence of fast, in-place, and ancilla-free quantum algorithms for integer multiplication between a quantum register and a fixed classical constant.

\section{Multiplication with a classical integer}\label{sec:classical multiplication}

\begin{theorem}[Modular multiplication by a classical integer mod $2^n$]\label{thm:classical mod mult}
    Let $z \in \Z_{2^n}$ and $a \in \Z_{2^n}^{\times}$. For any $\epsilon>0$, there exists a quantum circuit for performing the operation \begin{equation}
    \ket{z}_n \longmapsto \ket{az \bmod 2^n}_n
    \end{equation} that uses $O(n^{1 +\epsilon})$ gates, has depth $O(n\log n)$, and requires no ancilla qubits.
\end{theorem}
\begin{proof}
    Let $s=0$ (this implies $\alpha$ and $\beta$ are undefined and we use the convention $w^{\alpha} = w^\beta =1$). Let $t = \lceil n/2\rceil, m = \lfloor n/2\rfloor, 2^{m}x + y = z$, $b=c=a$ and apply Theorem \ref{thm:main theorem}.
\end{proof}

Before considering multiplication by an arbitrary classical integer, we introduce two new operations. For $a\in\mathbb Z_{2^m}$, let
\begin{equation}
\mathrm{tz}(a)=\log(\gcd(a,2^m)),
\end{equation}
denote the number of trailing zeroes in the binary representation of $a$. Note that $\mathrm{tz}(0)=m$. For a classically known integer $0\le j<n$, let $\mathbf{CS}_n(j)$ denote the cyclic shift
\begin{equation}
    \mathbf{CS}_n(j):
    \ket{w}_{n-j}\ket{x}_j
    \longmapsto
    \ket{x}_j\ket{w}_{n-j}.
\end{equation}
This moves the $j$ least significant qubits to the most significant position. 

\begin{remark}[Implementation of classical cyclic shifts]
\label{rem:classical cyclic shifts}
The operation $\mathbf{CS}_n(j)$ permutes $n$ qubits. Any permutation of $n$ positions can be written as a product of at most $n-1$ transpositions~\cite[\S~26.13]{NIST:DLMF}. Therefore, this operation can be implemented using at most $n-1$ SWAP gates, without ancilla qubits. Since $j$ is known classically, we can also perform this operation without applying any gates by simply relabelling qubits, provided subsequent operations use this new labelling.
\end{remark}
\begin{theorem}[Multiplication by a classical integer]\label{thm:classical full product}
    Let $z \in \Z_{2^n}, m = O(n)$ and $a \in \Z_{2^m}\setminus{\{0\}}$. For any $\epsilon>0$, there exists a quantum circuit for performing the operation \begin{equation}
    \ket{0}_{m}\ket{z}_n \longmapsto \ket{az}_{m+n}
    \end{equation} that uses $O(n^{1 +\epsilon})$ gates, has depth $O(n\log n)$, and requires no ancilla qubits.
\end{theorem}

\begin{proof}
    Let $j=\mathrm{tz}(a)$, which we compute classically, and let $a'=a/2^j$. Since $a'$ is odd, we can apply Theorem \ref{thm:main theorem} to $\ket{0}_{m-j}\ket{z}_n$, setting the parameters of the theorem as follows: $s=0, t =m-j,$ $m = n, x =0,$ $y =z, b =1$, and $c =a'$. This prepares $\ket{a'z}_{m+n-j}$. 
    We now apply $\mathbf{CS}_{m+n}(j)^{-1}$ to obtain
    \begin{flalign}
        &\xrightarrow{\mathbf{CS}_{m+n}(j)^{-1}}
        \ket{a'z}_{m+n-j}\ket{0}_j
        =
        \ket{az}_{m+n}.&&
    \end{flalign}

\end{proof}

\begin{remark}
    We can use a cheaper algorithm than the algorithm of Theorem \ref{thm:main theorem} to implement this multiplication. Namely, we can replace the first parametrised QFT in the algorithm of Theorem \ref{thm:main theorem} with a Hadamard transform, since $\qft{0}{m-j}{b}\ket{0}_{m-j} = H^{\otimes {m-j}}\ket{0}_{m-j}$. This change does not affect the asymptotic cost.
\end{remark}

\begin{remark}[Motivation for dirty qubit variants]\label{rem:dirty qubits motivation}
    In some algorithms, we may want to compute the product $az \in \Z_{2^{m+n}}$ to use it in subsequent computations, but be short on space to store this value. The next theorem considers working on top of idle qubits in a potentially unknown state. In the context of a larger algorithm, this operation may allow for even more space savings. 
    
    For example, if we wish to use some suitable function of $az$, we might pre-emptively perform a corrective operation with an unknown idle state, then build a combination of $az$ and this unknown state using the following construction, and then use this combination as $az$ would have been used. Reversing the multiplication then restores the idle register, with the net effect of the sequence being the intended effect of using the integer product alone. In this setting, the borrowed qubits are commonly referred to in the literature as dirty qubits.
\end{remark}

\begin{theorem}[Multiplication by a classical integer on mostly dirty qubits]\label{thm: classical dirty product}
    Let $z \in \Z_{2^n}$, $m =O(n)$ and $a \in \Z_{2^m}\setminus\{0\}$. Let $j=\mathrm{tz}(a)$. Let $x \in \Z_{2^{j}}$ and $y \in \Z_{2^{m-j}}$. For any $\epsilon>0$, there exists a quantum circuit for performing the operation 
    \begin{equation}
    \ket{0}_{1}\ket{x}_{j}\ket{y}_{m-j}\ket{z}_n \longmapsto \ket{2^{n+j}y+x + az}_{m+n+1}
    \end{equation} that uses $O(n^{1 +\epsilon})$ gates, has depth $O(n\log n)$, and requires no ancilla qubits.
\end{theorem}
\begin{proof}
As in Theorem~\ref{thm:classical full product}, compute $j=\mathrm{tz}(a)$ classically and let $a'=a/2^j$. Both $2^ny$ and $a'z$ can be represented using $m+n-j$ bits. However, their sum may require one extra bit. We therefore move the clean qubit $\ket{0}_1$ to be between the $x$- and $y$-registers, so that the multiplication we are trying to perform acts on $\ket{0}_1\ket{y}_{m-j}\ket{z}_n$, and leaves $\ket{x}_{j}$ unchanged. To do this, apply $\mathbf{CS}_m(j)^{-1}$ to the $x$- and $y$-registers, followed by $\mathbf{CS}_{m+1}(j)$ to the first $m+1$ qubits:
\begin{flalign}
    &\xrightarrow{\mathbf{CS}_m(j)^{-1}}
    \ket{0}_1\ket{y}_{m-j}\ket{x}_j\ket{z}_n, &&\\
    &\xrightarrow{\mathbf{CS}_{m+1}(j)}
    \ket{x}_j\ket{0}_1\ket{y}_{m-j}\ket{z}_n. &&
\end{flalign}
\emph{Since $j$ is known classically, the same operation can be performed by $\mathbf{CS}_{j+1}(j)$ on the leading clean qubit and the $x$-register. We retain the two shifts here to make the parallel with the quantum-integer construction in Theorem~\ref{thm:quantum dirty product} explicit.}

Let $s=0, t=m-j+1, m=n, x=y, y=z, b=1, c=a'$ and apply the circuit from Theorem~\ref{thm:main theorem} to $\ket{0}_1\ket{y}_{m-j}\ket{z}_n$. Since $2^ny+a'z<2^{m+n-j+1}$, this gives
\begin{flalign}
    &\xrightarrow{\text{Theorem~\ref{thm:main theorem}}}
    \ket{x}_j\ket{2^ny+a'z}_{m+n-j+1}. &&
\end{flalign}
We now apply $\mathbf{CS}_{m+n+1}(j)^{-1}$ to obtain
\begin{flalign}
    &\xrightarrow{\mathbf{CS}_{m+n+1}(j)^{-1}}
    \ket{2^ny+a'z}_{m+n-j+1}\ket{x}_j
    =
    \ket{2^{n+j}y+x+az}_{m+n+1}. &&
\end{flalign}
\end{proof}

It should be clear that Theorem \ref{thm:classical full product} is the special case of Theorem \ref{thm: classical dirty product} where $x=y=0$ and the overflow qubit $\ket{0}_1$ is omitted.

We show in the next section that our main theorem, Theorem \ref{thm:main theorem}, yields fast, in-place algorithms for multiplying quantum registers. All of these algorithms require at most a logarithmic number of clean ancillas, and some are completely ancilla-free.

\section{Multiplication with a quantum integer}\label{sec:quantum-int-mul}
\subsection{Subroutines}\label{subsec:quantum subroutines}
The preceding algorithms compute the trailing-zero count of an integer classically and use it to determine the required cyclic shifts. When the multiplier is a quantum integer, we must perform these operations coherently. We therefore introduce reversible evaluation of the trailing-zero function and cyclic shifts controlled by a quantum register. Quantum circuits for trailing-zero counting and controlled shifts have been studied previously \cite{haenerQuantumCircuitsFloatingPoint2018,mukhopadhyayImprovedQuantumCircuits2026}. We provide alternate constructions for these operations below.

Let $x\in\Z_{2^m}$ and $n\ge\lceil\log(m+1)\rceil$. Then,
\begin{align}
    \mathbf{TZ}_{m,n}:\ket{x}_m\ket{0}_n \longmapsto \ket{x}_m\ket{\mathrm{tz}(x)}_n.
\end{align}
This computes the trailing-zero count while preserving the input register. As usual, we use subscripts to indicate the sizes of the registers that we are applying the gate to. The first subscript always refers to the control register and the second to the target. For this operation to be well-defined,
$n\geq \lceil \log (m+1) \rceil$ .

\begin{restatable}[Trailing zero count]{lemma}{trailingzerocount}\label{lem:trailing zero circuit gate count}
    Let $m\ge1$. There exists a quantum circuit for $\mathbf{TZ}_{m,\lceil\log (m+1)\rceil}$ that uses $O(m)$ gates and no ancillas.
\end{restatable}
\begin{proof}
    The proof of this lemma is left for Appendix \ref{sec:appendix:additional}.
\end{proof}
Next, we consider the operation $\mathbf{CS}_{m,n}$ acting on $m+n$ qubits, defined as follows. Let $z\in\Z_{2^n}$ and let $y\in\Z_{2^m}$ such that $0\leq y<n$. Consider $x\in\Z_{2^y}$ and $w\in\Z_{2^{n-y}}$ such that $z=2^yw+x$. Then,
\begin{equation}
\begin{split}
    \mathbf{CS}_{m,n}:\ket{y}_m\ket{z}_n \longmapsto& \ket{y}_m\bigl(\mathbf{CS}_n(y)\ket{z}_n\bigr) \\
    =&\ket{y}_m\ket{x}_y\ket{w}_{n-y}.
\end{split}
\end{equation}
Controlled on the $y$-register, $\mathbf{CS}_{m,n}$ cyclically shifts the $y$ least significant qubits of the $z$-register, making them the most significant qubits. As with $\mathbf{TZ}_{m,n}$, the first subscript of $\mathbf{CS}_{m,n}$ always refers to the control register and the second to the target.
Observe that the action of $(\mathbf{CS}_{\lceil \log m \rceil,m}) \circ (\mathbf{I}_1 \otimes \mathbf{TZ}_{m-1,\lceil \log m \rceil})$ is well-defined on $\ket{x}_{m}\ket{0}_{\lceil \log m \rceil}$ for $x \in \Z_{2^m}\setminus\{0\}$. The composition of these operations prepares $\ket{0}_{\mathrm{tz}(x)}\ket{x/2^{\mathrm{tz}(x)}}_{m-{\mathrm{tz}(x)}}\ket{\mathrm{tz}(x)}_{\lceil \log m \rceil}$ since $\mathrm{tz}(x) = \mathrm{tz}(x_{m-2}\cdots x_{0})$.

\begin{restatable}[Cyclic shift]{lemma}{divisionbytwopower}
\label{lem:division two power circuit gate count}
Let $n\ge2$ and $1\le m\le\lceil\log n\rceil$.
There exists a quantum circuit for $\mathbf{CS}_{m,n}$ that uses $O(n\log n)$ gates and no ancillas.
\end{restatable}
\begin{proof}
Let $y=\sum_{i=0}^{m-1}2^i y_i$. We iterate over the bits of $y$ and apply $\mathbf{CS}_n(2^i)$ to the target register, controlled on $y_i$, giving the required total shift by $y$. By Remark~\ref{rem:classical cyclic shifts}, each iteration uses at most $n-1$ controlled SWAP gates. Therefore, we use at most $m(n-1)\le nm$ controlled SWAPs in total. Each controlled SWAP uses one Toffoli and two CNOTs, giving $O(nm)=O(n\log n)$ gates.
\end{proof}
\noindent With these operations and our main theorem in hand, we prove a number of results. 

\subsection{Main Results}
\begin{theorem}[Modular multiplication by a quantum integer mod $2^n$]\label{thm:quantum mod mult}
    Let $z \in \Z_{2^n}$. Let $m=O(n)$ and let $x \in \Z_{2^m}^{\times}$. For any $\epsilon>0$, there exists a quantum circuit for performing the operation \begin{equation}
        \ket{x}_m\ket{z}_n \longmapsto \ket{x}_m\ket{xz \bmod 2^n}_n
    \end{equation} that uses $O(n^{1 +\epsilon})$ gates and no ancilla qubits.
\end{theorem}
\begin{proof}
    Let $s=m, t = \lceil n/2\rceil, m = \lfloor n/2 \rfloor, \alpha = \beta =1,  w=x, 2^{m}x+y = z, b =c =1$ and apply Theorem \ref{thm:main theorem}.
\end{proof}
\begin{remark}
    It should be clear how to achieve $\ket{x}_{m}\ket{z}_n \mapsto \ket{x}_{m}\ket{axz \bmod 2^{n}}_n$ for some $a \in \Z \setminus 2\Z$. Simply let $b=c=a$ in the algorithm of Theorem \ref{thm:main theorem} instead. This shows that performing classical and quantum modular multiplication is possible within a single construction.
\end{remark}

\begin{theorem}[Multiplication by a quantum integer]\label{thm:quantum full product}
    Let $z \in \Z_{2^n}$, $m=O(n)$ and $x \in \Z_{2^m}\setminus\{0\}$. For any $\epsilon>0$, there exists a quantum circuit for performing the operation \begin{equation}
        \ket{x}_m\ket{0}_m\ket{z}_n \longmapsto \ket{x}_m\ket{xz}_{m+n}
    \end{equation} that uses $O(n^{1 +\epsilon})$ gates and requires $\lceil\log  m\rceil$ ancilla qubits. For the special case $x \in \Z_{2^m}^{\times}$ no ancilla qubits are required. 
\end{theorem}
\begin{proof}
    We provide an algorithm proving the claim. The main idea is the same as in Theorem~\ref{thm:classical full product}. We separate the odd factor $x/2^{\mathrm{tz}(x)}$, perform multiplication with this factor, and then rescale the result by $2^{\mathrm{tz}(x)}$.
    
    Since, $x \in \Z_{2^m}\setminus\{0\}$, $0\leq \mathrm{tz}(x) \leq m-1$. Note also that $\mathrm{tz}(x) = \mathrm{tz}(x_{m-2}\cdots x_{0})$. On input $\ket{x}_{m}\ket{0}_{\lceil \log m\rceil}\ket{0}_m\ket{z}_{n}$, apply the following operations
    \begin{flalign}
    &\xrightarrow{\mathbf{I}_1 \otimes \mathbf{TZ}_{m-1, \lceil \log m\rceil}}\ket{x_{m-1}}_1\ket{x_{m-2} \cdots x_{0}}_{m-1}\ket{\mathrm{tz}(x)}_{\lceil \log m\rceil}\ket{0}_{m}\ket{z}_{n}, &&
    \end{flalign}
    \begin{flalign}
    &\xrightarrow{\mathbf{CS}_{\lceil \log m\rceil, m}}\ket{x/2^{\mathrm{tz}(x)}}_{m}\ket{\mathrm{tz}(x)}_{\lceil \log m\rceil}\ket{0}_{m}\ket{z}_{n}. &&
    \end{flalign}
    Let $s=m, t=m, m=n, \alpha =0, \beta =1, w =x/2^{\mathrm{tz}(x)},  x = 0, y =z, b = c =1$ and apply Theorem \ref{thm:main theorem} to $\ket{x/2^{\mathrm{tz}(x)}}_{m}\ket{0}_{m}\ket{z}_{n}$
    \begin{flalign}\label{eq:quantum clean product middle equation}
    &\xrightarrow{\text{Theorem \ref{thm:main theorem}}} \ket{x/2^{\mathrm{tz}(x)}}_{m}\ket{\mathrm{tz}(x)}_{\lceil \log m\rceil}\ket{xz/2^{\mathrm{tz}(x)}}_{m+n}. &&
    \end{flalign}
    \begin{flalign}
    &\xrightarrow{\mathbf{CS}_{\lceil \log m\rceil, m+n}^{-1}} \ket{x/2^{\mathrm{tz}(x)}}_{m}\ket{\mathrm{tz}(x)}_{\lceil \log m\rceil}\ket{xz}_{m+n}. &&
    \end{flalign}
    \begin{flalign}
    &\xrightarrow{\mathbf{CS}_{\lceil \log m\rceil, m}^{-1}} \ket{x}_{m}\ket{\mathrm{tz}(x)}_{\lceil \log m\rceil}\ket{xz}_{m+n}. &&
    \end{flalign}
    \begin{flalign}
        &\xrightarrow{\mathbf{I}_1 \otimes \mathbf{TZ}_{m-1, \lceil \log m\rceil}^{-1}} \ket{x}_{m}\ket{0}_{\lceil \log m\rceil}\ket{xz}_{m+n}. &&
    \end{flalign}
    Since $m=O(n)$, by Theorem \ref{thm:main theorem} and Lemmas \ref{lem:trailing zero circuit gate count} and \ref{lem:division two power circuit gate count} this algorithm uses $O(n^{1+\epsilon})$ gates and $\lceil \log  m\rceil$ ancillas, for any $\epsilon>0$. For the special case $x \in \Z_{2^m}^{\times}$ the cyclic shifts and trailing zero computations are unnecessary, since $\mathrm{tz}(x) =0$. Theorem \ref{thm:main theorem} can be applied directly, requiring no ancilla qubits.
\end{proof}
\begin{remark}
    We can use a cheaper algorithm than the algorithm of Theorem \ref{thm:main theorem} to implement the multiplication taking place in Equation \ref{eq:quantum clean product middle equation}. Namely, we can replace the first parametrised QFT in the algorithm of Theorem \ref{thm:main theorem} with a Hadamard transform, since $\qft[\alpha]{s}{t}{b}\ket{w}_s\ket{0}_{t} = \ket{w}_sH^{\otimes {t}}\ket{0}_{t}$. This change does not affect the asymptotic cost. 
\end{remark}
\begin{remark}
    For $a \in \Z_{2^t}\setminus \{0\}$, $t=O(n)$, combining Theorems \ref{thm:classical full product} and \ref{thm:quantum full product} shows that the transformation $\ket{x}_m\ket{0}_{t+m}\ket{z}_n \mapsto \ket{x}_{m}\ket{axz}_{t+m+n}$ can be performed using $O(n^{1+\epsilon})$ gates and $\lceil \log  m\rceil$ ancillas. However, it should be clear that a cheaper algorithm, albeit with the same asymptotic cost, is available. First, compute $\mathrm{tz}(a)$ classically. Then, let $a' = a/2^{\mathrm{tz}(a)}$, and apply Theorem \ref{thm:quantum full product} to $\ket{x}_m\ket{0}_{t+m-\mathrm{tz}(a)}\ket{z}_n$ setting $c=a'$ instead. This will prepare $\ket{x}_{m}\ket{0}_{\mathrm{tz}(a)}\ket{xa'z}_{t+m+n-\mathrm{tz}(a)}$. Applying $\mathbf{CS}^{-1}_{t+m+n}(\mathrm{tz}(a))$, which by Remark \ref{rem:classical cyclic shifts} requires no additional gates, prepares $\ket{x}_{m}\ket{xaz}_{t+m+n}$. This shows that classical and quantum multiplication can be implemented in a single construction.
\end{remark}

Finally, we consider the scenario where we want to compute the product $xz \in \Z_{2^{m+n}}$ and we must work on top of dirty qubits. This provides a primitive for settings where qubits are available but initialising a large output register with clean ancillas is undesirable. See Remark \ref{rem:dirty qubits motivation} for further motivation.

\begin{theorem}[Multiplication by a quantum integer on mostly dirty qubits]\label{thm:quantum dirty product}
Let $z\in\Z_{2^n}$, $m=O(n)$ and $x\in\Z_{2^m}\setminus\{0\}$. For any $\epsilon>0$, there exists a quantum circuit for performing the operation
\begin{equation}
    \ket{x}_m\ket{0}_1\ket{w}_{\mathrm{tz}(x)}\ket{y}_{m-\mathrm{tz}(x)}\ket{z}_n \longmapsto \ket{x}_m\ket{2^{n+\mathrm{tz}(x)}y+w+xz}_{m+n+1}
\end{equation}
that uses $O(n^{1+\epsilon})$ gates and requires $\lceil\log m\rceil$ ancilla qubits. 
For the special case $x \in \Z_{2^m}^{\times}$ no ancilla qubits are required. The corresponding operation in this case is
\begin{equation}
    \ket{x}_m\ket{0}_1\ket{y}_{m}\ket{z}_n \longmapsto \ket{x}_m\ket{2^{n}y+xz}_{m+n+1}.
\end{equation}
\end{theorem}

\begin{proof}
We provide an algorithm similar to that of Theorem~\ref{thm:quantum full product}. We first transform the $x$ register and rearrange the target register so that we can perform the multiplication using the odd factor $x/2^{\mathrm{tz}(x)}$ of $x$. After this multiplication, we rescale the result by $2^{\mathrm{tz}(x)}$ and restore $x$.

On input $\ket{x}_m\ket{0}_{\lceil\log m\rceil}\ket{0}_1\ket{w}_{\mathrm{tz}(x)}\ket{y}_{m-\mathrm{tz}(x)}\ket{z}_n$, compute $\mathrm{tz}(x)$ from the least significant $m-1$ qubits of $x$,
\begin{flalign}
    &\mathord{\xrightarrow{\mathbf{I}_1 \otimes  \mathbf{TZ}_{m-1,\lceil\log m\rceil}\ }}\;\ket{x}_m\ket{\mathrm{tz}(x)}_{\lceil\log m\rceil}\ket{0}_1\ket{w}_{\mathrm{tz}(x)}\ket{y}_{m-\mathrm{tz}(x)}\ket{z}_n. &&
\end{flalign}
Next, apply $\mathbf{CS}_{\lceil\log m\rceil,m}$ to the $x$-register, controlled on $\mathrm{tz}(x)$. All applications of $\mathbf{CS}$ in this algorithm are controlled on $\mathrm{tz}(x)$.
\begin{flalign}
    &\mathord{\xrightarrow{\ \mathbf{CS}_{\lceil\log m\rceil,m}\ }}\;\ket{x/2^{\mathrm{tz}(x)}}_m\ket{\mathrm{tz}(x)}_{\lceil\log m\rceil}\ket{0}_1\ket{w}_{\mathrm{tz}(x)}\ket{y}_{m-\mathrm{tz}(x)}\ket{z}_n. &&
\end{flalign}
We now need to place the clean qubit between $w$ and $y$, to accommodate the carry from $2^ny+xz/2^{\mathrm{tz}(x)}$. To do this, first apply $\mathbf{CS}_{\lceil\log m\rceil,m}^{-1}$ to the $m$ qubits $\ket{w}_{\mathrm{tz}(x)}\ket{y}_{m- \mathrm{tz}(x)}$,
\begin{flalign}
    &\mathord{\xrightarrow{\ \mathbf{CS}_{\lceil\log m\rceil,m}^{-1}\ }}\;\ket{x/2^{\mathrm{tz}(x)}}_m\ket{\mathrm{tz}(x)}_{\lceil\log m\rceil}\ket{0}_1\ket{y}_{m-\mathrm{tz}(x)}\ket{w}_{\mathrm{tz}(x)}\ket{z}_n. &&
\end{flalign}
Then, apply $\mathbf{CS}_{\lceil\log m\rceil,m+1}$ to the $m+1$ qubits $\ket{0}_1\ket{y}_{m -\mathrm{tz}(x)}\ket{w}_{\mathrm{tz}(x)}$,
\begin{flalign}
    &\mathord{\xrightarrow{\ \mathbf{CS}_{\lceil\log m\rceil,m+1}\ }}\;\ket{x/2^{\mathrm{tz}(x)}}_m\ket{\mathrm{tz}(x)}_{\lceil\log m\rceil}\ket{w}_{\mathrm{tz}(x)}\ket{0}_1\ket{y}_{m-\mathrm{tz}(x)}\ket{z}_n. &&
\end{flalign}
Note, for $\mathrm{tz}(x)=0$, both shifts are the identity and the register $w$ has zero qubits. We can write this state as
\begin{flalign}
    &=\ket{x/2^{\mathrm{tz}(x)}}_m\ket{\mathrm{tz}(x)}_{\lceil\log m\rceil}\ket{2^{m-\mathrm{tz}(x)+1}w+y}_{m+1}\ket{z}_n. &&
\end{flalign}
Since $x/2^{\mathrm{tz}(x)}$ is odd, we can now apply Theorem~\ref{thm:main theorem} (with $s=m$, $t=m+1$, $m=n$, $\alpha=0$, $\beta=1$ and $b=c=1$).
\begin{flalign}
    &\mathord{\xrightarrow{\ \text{Theorem~\ref{thm:main theorem}}\ }}\;\ket{x/2^{\mathrm{tz}(x)}}_m\ket{\mathrm{tz}(x)}_{\lceil\log m\rceil}\ket{2^{m+n-\mathrm{tz}(x)+1}w+2^ny+xz/2^{\mathrm{tz}(x)}}_{m+n+1}. &&
\end{flalign}
Since $y,x/2^{\mathrm{tz}(x)}<2^{m-\mathrm{tz}(x)}$ and $z<2^n$, we have
\begin{align}
    0\le 2^ny+xz/2^{\mathrm{tz}(x)}<2^{m+n-\mathrm{tz}(x)+1}.
\end{align}
Therefore, no carry propagates to $w$, and the state is
\begin{flalign}
    &=\ket{x/2^{\mathrm{tz}(x)}}_m\ket{\mathrm{tz}(x)}_{\lceil\log m\rceil}\ket{w}_{\mathrm{tz}(x)}\ket{2^ny+xz/2^{\mathrm{tz}(x)}}_{m+n-\mathrm{tz}(x)+1}. &&
\end{flalign}
We now apply $\mathbf{CS}_{\lceil\log m\rceil,m+n+1}^{-1}$ to the last $m+n+1$ qubits, moving $w$ to the least significant positions.
\begin{flalign}
    &\mathord{\xrightarrow{\ \mathbf{CS}_{\lceil\log m\rceil,m+n+1}^{-1}\ }}\;\ket{x/2^{\mathrm{tz}(x)}}_m\ket{\mathrm{tz}(x)}_{\lceil\log m\rceil}\ket{2^ny+xz/2^{\mathrm{tz}(x)}}_{m+n-\mathrm{tz}(x)+1}\ket{w}_{\mathrm{tz}(x)} &&\\
    &=\ket{x/2^{\mathrm{tz}(x)}}_m\ket{\mathrm{tz}(x)}_{\lceil\log m\rceil}\ket{2^{n+\mathrm{tz}(x)}y+xz+w}_{m+n+1}. &&
\end{flalign}
Finally, we restore $x$ by applying $\mathbf{CS}_{\lceil\log m\rceil,m}^{-1}$ to the first $m$ qubits, and invert the trailing-zero computation.
\begin{flalign}
    &\mathord{\xrightarrow{\ \mathbf{CS}_{\lceil\log m\rceil,m}^{-1}\ }}\;\ket{x}_m\ket{\mathrm{tz}(x)}_{\lceil\log m\rceil}\ket{2^{n+\mathrm{tz}(x)}y+xz+w}_{m+n+1}, &&\\
    &\mathord{\xrightarrow{\mathbf{I}_1 \otimes \mathbf{TZ}_{m-1,\lceil\log m\rceil}^{-1}\ }}\;\ket{x}_m\ket{0}_{\lceil\log m\rceil}\ket{2^{n+\mathrm{tz}(x)}y+xz+w}_{m+n+1}. &&
\end{flalign}
Since $m=O(n)$, the trailing-zero computations and cyclic shifts use $O(n\log n)$ gates by Lemmas~\ref{lem:trailing zero circuit gate count} and~\ref{lem:division two power circuit gate count}. Together with Theorem~\ref{thm:main theorem}, this gives $O(n^{1+\epsilon})$ gates and $\lceil\log m\rceil$ ancilla qubits. For the special case $x \in \Z_{2^m}^{\times}$ the cyclic shifts and trailing zero computations are unnecessary, since $\mathrm{tz}(x) =0$ and there is no $w$-register. In this case, Theorem \ref{thm:main theorem} can be applied directly, requiring no ancilla qubits.
\end{proof}

\begin{remark}\label{rem:zero-input-case}
In Theorems~\ref{thm:quantum full product} and~\ref{thm:quantum dirty product}, we restrict $x$ to be non-zero so that the stated transformations are unitary. If $x=0$ is allowed, the appropriate behaviour depends on the surrounding algorithm that the multiplication sits within. As an example, replacing Theorem~\ref{thm:main theorem} with Theorem~\ref{thm:main theorem extended} in the above proofs gives an extension to all $x$, with the same gate and ancilla bounds and unchanged behaviour for $x\ne0$.
\end{remark}

In the next section, we use our modular multiplication circuit to develop in-place algorithms for computing the multiplicative inverse of a quantum integer modulo a power of two. We prove that these algorithms attain the same asymptotic upper bound on gate count as quantum modular multiplication, even without ancilla qubits.

\section{Multiplicative inverse}\label{sec:inversion}
In this section, we use the symbols $\times$ and $\div$ to represent quantum circuits that perform modular multiplication with a quantum multiplier. In more detail, these operations are defined for $x\in \Z_{2^m}^{\times}$ and $y \in \Z_{2^n}$ as follows,
\begin{align*}
    &\times(x):\ket{x}_{m}\ket{y}_n \longmapsto \ket{x}_{m}\ket{yx \bmod 2^n}_{n}, &&\div(x): \ket{x}_{m}\ket{y}_{n} \longmapsto \ket{x}_{m}\ket{y(x^{-1}) \bmod 2^n}_n.
\end{align*}
All arithmetic on \(n\) qubit registers is understood to be mod \(2^n\), so from now on, we drop the modulus for ease of notation.

For $m=O(n)$, by Theorem \ref{thm:quantum mod mult}, the operation $\times$ can be performed using $O(n^{1+\epsilon})$ gates and no ancilla qubits, for any $\epsilon>0$. $\div$ is the inverse operation. We use these costs when justifying the main results of this section.

Before proceeding to these results, we prove the following lemma.
\begin{lemma}[Computing a partial inverse]\label{lem:partial-inverse}
Let $1\leq n \leq m$. Let $x\in\Z_{2^m}^{\times}$ and let $x^{-1}$ denote the multiplicative inverse of $x$ in $\Z_{2^m}^{\times}$. For any $\epsilon>0$, there exists a quantum circuit for performing the operation
\begin{equation}\label{eq:partial-inverse}
 \ket{x}_m\ket{0}_n\longmapsto
 \ket{x}_m\ket{x^{-1}\bmod2^n}_n
\end{equation}
that uses $O(n^{1+\epsilon})$ gates and no ancilla qubits.
\end{lemma}
\begin{proof}
    Consider $x_{\mathrm{lo}}, (x^{-1})_{\mathrm{lo}} \in \Z_{2^n}$ and $x_{\mathrm{hi}}, (x^{-1})_{\mathrm{hi}} \in \Z_{2^{m-n}}$ such that \begin{equation}
     x=2^n x_{\rm hi}+x_{\rm lo},\qquad
     x^{-1}=2^n(x^{-1})_{\rm hi}+(x^{-1})_{\rm lo}.
    \end{equation}
    Expanding the product $xx^{-1}$ gives
    \begin{equation}
        \begin{split}
          2^{2n}x_{\rm hi}(x^{-1})_{\rm hi}
         +2^n\bigl(x_{\rm hi}(x^{-1})_{\rm lo}
         +x_{\rm lo}(x^{-1})_{\rm hi}\bigr)
         +x_{\rm lo}(x^{-1})_{\rm lo}
         \equiv1\pmod{2^m}.
        \end{split}
    \end{equation}
    Reducing modulo $2^n$ gives $x_{\rm lo}(x^{-1})_{\rm lo}\equiv1\pmod{2^n}$. Since $x_\mathrm{lo}$ has a unique inverse in $\Z_{2^n}^{\times}$ and $(x^{-1})_{\mathrm{lo}} \in \Z_{2^n}$ we know that $(x_{\mathrm{lo}})^{-1} = (x^{-1})_{\mathrm{lo}}$.
    This observation reveals an algorithm proving the claim. On input $\ket{x}_m\ket{0}_n$, apply
    \begin{flalign}
    &\xrightarrow{\mathbf{I}_{m+n-1} \otimes X} \ket{x_{\rm hi}}_{m-n}\ket{x_{\rm lo}}_n\ket{1}_n,&&
    \end{flalign}
    \begin{flalign}
    &\xrightarrow{\div(x_{\rm lo})} \ket{x_{\rm hi}}_{m-n}\ket{x_{\rm lo}}_n\ket{(x^{-1})_{\rm lo}}_n.&&
    \end{flalign}
\end{proof}

\begin{theorem}[In-place modular inversion]\label{thm:modular-inv}
Let $m\ge2$ and $n=\lceil m/2\rceil$. Let $x\in\Z_{2^m}^{\times}$ and let $x^{-1}$ denote the multiplicative inverse of $x$ in $\Z_{2^m}^{\times}$.
For any $\epsilon>0$, there exists a quantum circuit for performing the operation
\begin{equation}\label{eq:half-inverse}
 \ket{x}_m\longmapsto
 \ket{x^{-1}}_m
\end{equation}
that uses $O(m^{1+\epsilon})$ gates and requires $n$ ancilla qubits.
\end{theorem}
\begin{proof}
As in Lemma \ref{lem:partial-inverse}, consider $x_{\mathrm{lo}}, (x^{-1})_{\mathrm{lo}} \in \Z_{2^n}$ and $x_{\mathrm{hi}},  (x^{-1})_{\mathrm{hi}} \in \Z_{2^{m-n}}$ such that  \begin{equation}
     x=2^n x_{\rm hi}+x_{\rm lo},\qquad
     x^{-1}=2^n(x^{-1})_{\rm hi}+(x^{-1})_{\rm lo}.
    \end{equation}
We showed in Lemma \ref{lem:partial-inverse} that $(x^{-1})_{\rm lo}$ equals $(x_{\rm lo})^{-1}$, the multiplicative inverse of $x_{\rm lo}$ in $\Z_{2^n}^{\times}$. In particular, this implies that $x_{\rm lo}(x^{-1})_{\rm lo} = 2^n c + 1$ for some $c \in \Z_{2^n}$. With these observations in hand,
we provide an algorithm proving the claim. On input $\ket{x}_m\ket{0}_n$, apply
\begin{flalign}
    &\xrightarrow{\text{Lemma} \ \ref{lem:partial-inverse}} \ket{x}_{m}\ket{(x^{-1})_{\rm lo}}_{n}. &&
\end{flalign}
Then multiply this partial inverse into $\ket{x}_m$
\begin{flalign}
    & \xrightarrow{\times((x^{-1})_{\rm lo})} \ket{x(x^{-1})_{\rm lo}}_{m}\ket{(x^{-1})_{\rm lo}}_{n} = \ket{x_{\rm hi}(x^{-1})_{\rm lo}+c}_{m-n}\ket{1}_n\ket{(x^{-1})_{\rm lo}}_n, &&
\end{flalign}
We can reset the middle $n$ qubit register to $\ket{0}_n$ with an $X$ gate and ignore it for the rest of the algorithm. The state is thus,
\begin{align}\label{eq:inversion pause equation}
    \ket{x_{\rm hi}(x^{-1})_{\rm lo}+c}_{m-n} \ket{(x^{-1})_{\rm lo}}_n.
\end{align}
Finally, multiply the $m-n$ qubit register with $-(x^{-1})_{\rm lo}$, 
\begin{flalign}
    & \xrightarrow{\times(-(x^{-1})_{\rm lo})} \ket{-2^nx_{\rm hi}(x^{-1})_{\rm lo}^{2} -2^{n}c (x^{-1})_{\rm lo} +(x^{-1})_{\rm lo}}_{m}, &&
\end{flalign}
since $2^n (y \bmod 2^{m-n}) = 2^{n}y \bmod 2^m$. 
The result follows from
\begin{equation}
\begin{aligned}
&x\bigl(-2^nx_{\rm hi}(x^{-1})_{\rm lo}^{2}
-2^nc(x^{-1})_{\rm lo}
+(x^{-1})_{\rm lo}\bigr)\\
&\quad=(2^nx_{\rm hi}+x_{\rm lo})
\bigl(-2^nx_{\rm hi}(x^{-1})_{\rm lo}^{2}
-2^nc(x^{-1})_{\rm lo}
+(x^{-1})_{\rm lo}\bigr)\\
&\quad\equiv 2^nx_{\rm hi}(x^{-1})_{\rm lo}
-2^nx_{\rm hi}(x^{-1})_{\rm lo}
+(-2^nc+1)x_{\rm lo}(x^{-1})_{\rm lo}
\pmod{2^m}\\
&\quad\equiv (-2^nc+1)(2^nc+1)
\equiv 1\pmod{2^m}.
\end{aligned}
\end{equation}
\end{proof}

\subsection{Ancilla-efficient inversion}
Finally, we present an in-place modular inversion algorithm for the more general setting where we have $0\leq s \leq m-1$ ancillas at our disposal. We show that, with access to any number of ancillas in this range, our algorithm retains the fast asymptotic cost of quantum modular multiplication. This remains true even in the absence of ancillas.

In Theorem \ref{thm:modular-inv}, notice that the state of the algorithm at Equation \ref{eq:inversion pause equation} is $\ket{z}_{m-n}\ket{y}_{n}$ for integers $y, z$ satisfying the relation $xy \equiv 1 + 2^n z \pmod{2^m}$. We showed in this theorem that, if $n= \lceil m/2 \rceil$, the only operation that remains to prepare $\ket{x^{-1}}_m$ from $\ket{z}_{m-n}\ket{y}_n$ is a multiplication by $-y$ on the $z$-register. 

In Theorem \ref{thm:budget} below, we will see that, for any $n\geq \lceil m/2 \rceil$, multiplying the $z$-register by $-y$ always prepares $\ket{x^{-1}}_{m}$, as long as the integers $y$ and $z$ satisfy the previous relation. This motivates the following lemma.
\begin{restatable}[Doubling the number of inverse bits]{lemma}{InversionDouble}\label{lem:double}
Let $2 \leq 2h < m$. Let $x\in\Z_{2^m}^{\times},y \in \Z_{2^h}^{\times}$ and $z \in \Z_{2^{m-h}}$ such that $xy \equiv 1 + 2^h z \pmod{2^m}$. There exists a quantum circuit for performing the operation
\begin{equation}
 \ket z_{m-h}\ket y_h\longmapsto
 \ket{z'}_{m-2h}\ket{y'}_{2h},
\end{equation}
with $y' \in \Z_{2^{2h}}, z' \in \Z_{2^{m-2h}}$ satisfying the relation $xy'\equiv1+2^{2h}z'\pmod{2^m}$. The circuit uses
six modular multiplications, four modular additions and no ancilla qubits.
\end{restatable}
\begin{proof} 
Our algorithm for performing this inverse-doubling operation is derived from the classical Newton--Hensel iteration~\cite[Section~2.5]{brentzimmermann2010modern}. The proof is left for Appendix \ref{sec:appendix:inversion}.
\end{proof}

\noindent With this lemma in hand, we prove our ancilla-efficient modular inversion algorithm.

\begin{theorem}[Ancilla-efficient in-place modular inversion]\label{thm:budget}
Let $m\ge2$, and $0\le s\le m-1$. Let $x\in\Z_{2^m}^{\times}$ and let $x^{-1}$ denote the multiplicative inverse of $x$ in $\Z_{2^m}^{\times}$.
For any $\epsilon>0$, there exists a quantum circuit for performing the operation
\begin{equation}
 \ket{x}_m\longmapsto
 \ket{x^{-1}}_m
\end{equation}
that uses $O(m^{1+\epsilon})$ gates and $s$ ancilla qubits. The circuit uses
$O\bigl(1+\log(m/(s+1))\bigr)$ modular multiplications and additions.
\end{theorem}
\begin{proof}
We present an algorithm proving the claim. The first step of the algorithm is to transform the input register $\ket{x}_m$ to $\ket{z}_{m-h}\ket{y}_{h}$ for some $z \in \Z_{2^{m-h}}, y \in \Z_{2^h}$ satisfying $xy \equiv 1+ 2^{h}z \pmod{2^{m}}$. After this is achieved, we can iteratively apply Lemma \ref{lem:double}. We stop when the size of the register that contains the partial inverse, denoted $h'$, satisfies $2h' \geq m$. At this point we have $\ket{z'}_{m-h'}\ket{y'}_{h'}$ for some $z' \in \Z_{2^{m-h'}}, y' \in \Z_{2^{h'}}$ satisfying $xy' \equiv 1+ 2^{h'}z' \pmod{2^{m}}$. We then apply,
\begin{flalign}
    &\xrightarrow{\times(-y')} \ket{-y'z'}_{m-h'}\ket{y'}_{h'}.&&
\end{flalign}
It follows that we have prepared $\ket{x^{-1}}_{m}$ since
\begin{equation}
    x(-2^{h'}y'z' + y') = xy'(-2^{h'}z' +1) \equiv (2^{h'}z'+1)(-2^{h'}z' +1) \equiv 1\pmod{2^m}.
\end{equation}

All that remains is to show that we can transform $\ket{x}_m$ to $\ket{z}_{m-h}\ket{y}_{h}$ with $xy \equiv 1+ 2^{h}z \pmod{2^{m}}$. For $s\geq 1$, we repeat the approach presented in Theorem \ref{thm:modular-inv}. On input $\ket{x}_m\ket{0}_s$, apply
\begin{flalign}
     &\xrightarrow{\text{Lemma } \ref{lem:partial-inverse}}\ket{x}_m\ket{x^{-1} \bmod 2^{s}}_{s}, &&
\end{flalign}
\begin{flalign}
     &\xrightarrow{\times(x^{-1} \bmod 2^s)}\ket{z}_{m-s}\ket{1}_{s}\ket{x^{-1} \bmod 2^{s}}_{s}. &&
\end{flalign}
Here the integer $z \in \Z_{2^{m-s}}$ clearly satisfies the relation $xy \equiv 1+2^{s}z \pmod{2^m}$ with $y = x^{-1} \bmod 2^s$. Thus, we have the conditions necessary to apply Lemma \ref{lem:double}, with $h=s$ to $\ket{z}_{m-s}\ket{x^{-1} \bmod 2^s}_s$. For $s=0$, the original input register $\ket{x}_m$ already has the desired form with $h=1$, $y=1$ and $z=(x-1)/2$.

Since $h$ initially equals $\max\{1,s\}\ge(s+1)/2$,
the total number of multiplications and additions is
$O\bigl(1+\log(m/(s+1))\bigr)$.
For any $\epsilon >0$, we can pick $0<\delta< \epsilon$. Each modular multiplication in the above algorithm can then be performed using $O(m^{1+\delta})$ gates. Thus, for any $0\leq s\leq m-1$ and any $\epsilon >0$, the total algorithm requires $O\bigl(m^{1+\delta}(1+\log m)\bigr)=O(m^{1+\epsilon})$
gates and $s$ ancilla qubits.
\end{proof}

\begin{remark}\label{rem:kickstart-inversion}
    For $m\leq3$, every odd integer is its own multiplicative inverse modulo $2^m$. For $m>3$, write $x=8u+y$, where $y\in\{1,3,5,7\}$. Updating $u$ to $z=(uy+(y^2-1)/8)\bmod2^{m-3}$ gives $xy\equiv1+8z\pmod{2^m}$, allowing us to start the zero-ancilla algorithm with $h=3$. This uses one modular multiplication and $O(1)$ controlled additions, since $(y^2-1)/8=y_1+3y_2+2y_1y_2$ for $y=1+2y_1+4y_2$; the existing bit $y_0=1$ can be borrowed and restored. Thus, including this initialization, we save a few multiplications without changing the asymptotic cost of the algorithm.
\end{remark}

\begin{figure}[!htbp]
\centering
\includegraphics[width=\textwidth]{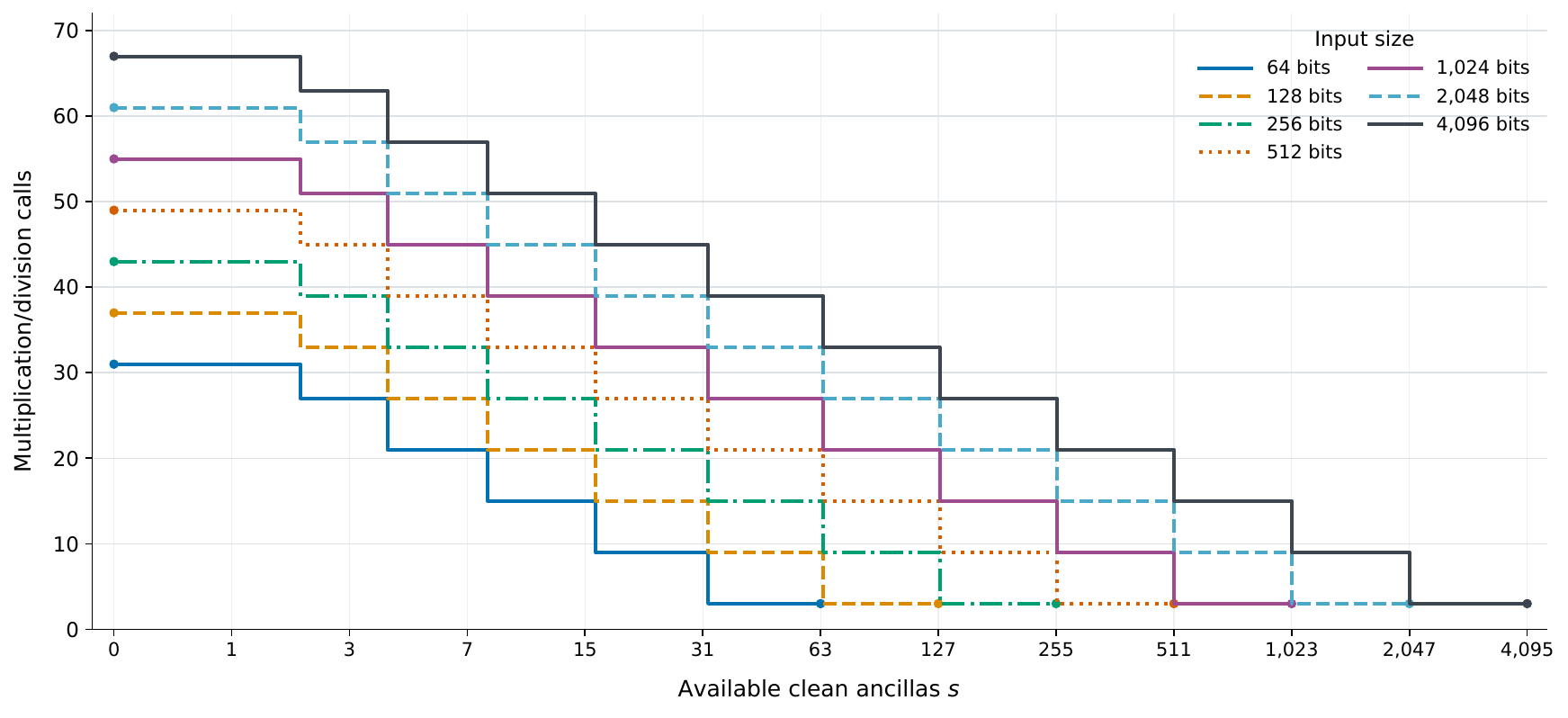}
\caption{Number of modular multiplications required for in-place inversion $\bmod \ 2^m$ as a function of number of clean ancillas available ($s$). Horizontal spacing is logarithmic in $s+1$. The multiplications do not take place on registers of the same size, so are not equivalent in terms of true gate count. \emph{Note: We do not take into account the optimizations from Remark~\ref{rem:kickstart-inversion} in this plot.}}
\label{fig:inverse-ancillas}
\end{figure}

Figure~\ref{fig:inverse-ancillas} illustrates the relationship between clean ancillas and modular multiplications for the algorithm proposed in Theorem \ref{thm:budget}. As an example, at $m=256$, allowing at most $0$, $8$, $32$, and $128$ clean ancillas gives $43$, $27$, $15$, and $3$ modular multiplications, respectively.

\section{Discussion}\label{sec:discussion}
In this work, we have presented several quantum algorithms for in-place multiplication and inversion that simultaneously achieve subquadratic gate counts and zero ancilla qubits (with at most a logarithmic number of ancillas required for multiplication by a general non-zero quantum integer). Our constructions build on the phase-encoded multiplication algorithms of Kahanamoku-Meyer and Yao~\cite{kahanamoku-meyerFastQuantumInteger2024} and the fast quantum Fourier transform construction of Cleve and Watrous~\cite{cleveFastParallelCircuits2000}. We show that a fast controlled and parametrised quantum Fourier transform leads to several in-place and ancilla-free multiplication algorithms arising naturally from a single unified theorem (Theorem~\ref{thm:main theorem}). We envision that this unified construction may also make our algorithms easier to code up, with the different variants obtained by toggling relevant parameters.

The gate counts we report are based on a family of recursive phase-encoded multiplication algorithms, inspired by classical $k$-way Toom--Cook multiplication, whose asymptotic gate counts can be made arbitrarily close to linear. However, the constants hidden in these bounds depend on the chosen recursion parameters. Identifying practical fault-tolerant instantiations therefore requires explicit resource estimates and a suitable choice of recursive decompositions and base cases. Our results also assume the exact implementation of arbitrary rotation gates. Resource estimates for a finite gate set must additionally account for the required approximation precision and the cost of synthesising these rotations.

Our constructions provide flexibility when arithmetic is used within a larger algorithm. The dirty register variants reduce the need for initially clean qubits, provided that the surrounding computation supports the required correction and restoration of the borrowed register. For modular inversion, additional clean ancillas reduce the number of multiplication and addition operations while retaining the same asymptotic gate costs. Since these operations act on registers of different sizes, their count alone does not determine the total gate cost. These choices should therefore be assessed in the context of a full algorithm and its available workspace.

Several future directions and open questions remain. An important next step would be to try and improve the depth bounds for multiplication by a quantum integer while retaining the gate and ancilla bounds obtained here. Another direction is to determine whether our modulo powers of two constructions can be extended to arbitrary moduli while retaining the same resource costs. Together with explicit finite-size resource estimates, these developments would clarify the range of applications for the methods presented in this work.

\section{Acknowledgements}
We thank Alex Cojocaru for helpful discussions and feedback on this work. D.~H and P.~W. were supported by the Ofgem Strategic Innovation Fund (SIF), delivered in partnership with Innovate UK (UKRI), under the `Network Security in a Quantum Future (NSiaQF)' project (Grant Reference: 10166255) and
by the EPSRC-funded `Quantum Advantage Pathfinder (QAP)' programme (Grant Reference: EP/X026167/1). A.~S. was supported by the Engineering and Physical Sciences Research Council (EPSRC) doctoral studentship (Grant Reference: EP/W524384/1). P.~W. was also supported by the EPSRC-funded `Quantum Advantage Turbocharger (QATCH)' programme (Grant Reference: UKRI4257) and the QCi3 Hub (Grant Reference: EP/Z53318X/1)

We used GPT-6 Astra and GPT-6.1 Sol to search the literature, proofread earlier drafts, and suggest minor textual revisions. We also used it to suggest proof techniques for Lemmas~\ref{lem:newPgatedepth} and~\ref{lem:double}. The authors reviewed and revised the AI-assisted material and take full responsibility for the content of this work.

\bibliographystyle{unsrt}
\bibliography{bib}

@misc{NIST:DLMF,
  key          = {{\relax DLMF}},
  title        = {{NIST Digital Library of Mathematical Functions}},
  howpublished = {\url{https://dlmf.nist.gov/}, Release 1.2.8 of 2026-09-15},
  url          = {https://dlmf.nist.gov/26.13},
  note         = {F.~W.~J. Olver, A.~B. {Olde Daalhuis}, D.~W. Lozier,
                  B.~I. Schneider, R.~F. Boisvert, C.~W. Clark,
                  B.~R. Miller, B.~V. Saunders, H.~S. Cohl,
                  and M.~A. McClain, eds.}
}

@article{rines2018high,
  title={High performance quantum modular multipliers},
  author={Rines, Rich and Chuang, Isaac},
  journal={arXiv preprint arXiv:1801.01081},
  year={2018}
}

@misc{gidney2017simple,
  author       = {Gidney, Craig},
  title        = {Simple Algorithm for Multiplicative Inverses mod $2^n$},
  howpublished = {https://algassert.com/post/1709},
  year         = {2017},
  month        = jun,
}

@article{kowada2006reversible,
  title = {Reversible Karatsuba's Algorithm.},
  author = {Kowada, Luis and Portugal, Renato and Figueiredo, Celina},
  year = 2006,
  month = jan,
  journal = {J. UCS},
  volume = {12},
  pages = {499--511}
}

@article{nieQuantumCircuitDesign2023,
  title = {Quantum {{Circuit Design}} for {{Integer Multiplication Based}} on {{Sch\"onhage}}--{{Strassen Algorithm}}},
  author = {Nie, Junhong and Zhu, Qinlin and Li, Meng and Sun, Xiaoming},
  year = 2023,
  month = dec,
  journal = {IEEE Transactions on Computer-Aided Design of Integrated Circuits and Systems},
  volume = {42},
  number = {12},
  pages = {4791--4802},
  issn = {1937-4151},
  doi = {10.1109/TCAD.2023.3279300},
  urldate = {2026-11-17}
}

@article{duttaQuantumCircuitsToomCook2018,
  title = {Quantum Circuits for {{Toom-Cook}} Multiplication},
  author = {Dutta, Srijit and Bhattacharjee, Debjyoti and Chattopadhyay, Anupam},
  year = 2018,
  month = jul,
  journal = {Physical Review A},
  volume = {98},
  number = {1},
  pages = {012311},
  issn = {2469-9926, 2469-9934},
  doi = {10.1103/PhysRevA.98.012311},
  urldate = {2026-11-17},
  langid = {english}
}

@article{parentImprovedReversibleQuantum2017,
  title={Improved reversible and quantum circuits for Karatsuba-based integer multiplication},
  author={Parent, Alex and Roetteler, Martin and Mosca, Michele},
  journal={arXiv preprint arXiv:1706.03419},
  year={2017},
  url={https://arxiv.org/abs/1706.03419}
}

@article{zalkaFastVersionsShors1998,
  title={Fast versions of Shor's quantum factoring algorithm},
  author={Zalka, Christof},
  journal={arXiv preprint quant-ph/9806084},
  year={1998}
}

@article{Bennett1973LogicalReversibility,
  author    = {Charles H. Bennett},
  title     = {Logical Reversibility of Computation},
  journal   = {IBM Journal of Research and Development},
  volume    = {17},
  number    = {6},
  pages     = {525--532},
  year      = {1973}
}

@article{Shor_1997,
   title={Polynomial-Time Algorithms for Prime Factorization and Discrete Logarithms on a Quantum Computer},
   volume={26},
   ISSN={1095-7111},
   url={http://dx.doi.org/10.1137/S0097539795293172},
   DOI={10.1137/s0097539795293172},
   number={5},
   journal={SIAM Journal on Computing},
   publisher={Society for Industrial & Applied Mathematics (SIAM)},
   author={Shor, Peter W.},
   year={1997},
   month=Oct, pages={1484–1509} }

@article{hallgren2007,
  title={Polynomial-time quantum algorithms for Pell's equation and the principal ideal problem},
  author={Hallgren, Sean},
  journal={Journal of the ACM (JACM)},
  volume={54},
  number={1},
  pages={1--19},
  year={2007},
  publisher={ACM, New York, NY, USA}
}

@article{regev2024efficientquantumfactoringalgorithm,
  title={An efficient quantum factoring algorithm},
  author={Regev, Oded},
  journal={Journal of the ACM},
  volume={72},
  number={1},
  pages={1--13},
  year={2025},
  publisher={ACM, New York, NY}
}

@article{Berry:2009cwy,
    author = "Berry, Dominic W. and Childs, Andrew M.",
    title = "{Black-box Hamiltonian simulation and unitary implementation}",
    eprint = "0910.4157",
    archivePrefix = "arXiv",
    primaryClass = "quant-ph",
    doi = "10.26421/QIC12.1-2-4",
    journal = "Quant. Inf. Comput.",
    volume = "12",
    number = "1-2",
    pages = "0029--0062",
    year = "2012"
}

@inproceedings{Ambainis_2011,
   title={Symmetry-Assisted Adversaries for Quantum State Generation},
   url={http://dx.doi.org/10.1109/CCC.2011.24},
   DOI={10.1109/ccc.2011.24},
   booktitle={2011 IEEE 26th Annual Conference on Computational Complexity},
   publisher={IEEE},
   author={Ambainis, Andris and Magnin, Loïck and Roetteler, Martin and Roland, Jeremie},
   year={2011},
   pages={167–177} }

@article{Jordan_2025,
   title={Optimization by decoded quantum interferometry},
   volume={646},
   ISSN={1476-4687},
   url={http://dx.doi.org/10.1038/s41586-025-09527-5},
   DOI={10.1038/s41586-025-09527-5},
   number={8086},
   journal={Nature},
   publisher={Springer Science and Business Media LLC},
   author={Jordan, Stephen P. and Shutty, Noah and Wootters, Mary and Zalcman, Adam and Schmidhuber, Alexander and King, Robbie and Isakov, Sergei V. and Khattar, Tanuj and Babbush, Ryan},
   year={2025},
   month=Oct, pages={831–836} }

@article{proos2004shorsdiscretelogarithmquantum,
  title={Shor's discrete logarithm quantum algorithm for elliptic curves},
  author={Proos, John and Zalka, Christof},
  journal={arXiv preprint quant-ph/0301141},
  year={2003}
}

@article{Kassal_2008,
   title={Polynomial-time quantum algorithm for the simulation of chemical dynamics},
   volume={105},
   ISSN={1091-6490},
   url={http://dx.doi.org/10.1073/pnas.0808245105},
   DOI={10.1073/pnas.0808245105},
   number={48},
   journal={Proceedings of the National Academy of Sciences},
   publisher={National Academy of Sciences},
   author={Kassal, Ivan and Jordan, Stephen P. and Love, Peter J. and Mohseni, Masoud and Aspuru-Guzik, Alán},
   year={2008},
   month=Dec, pages={18681–18686} }

@article{Babbush_2019,
   title={Quantum simulation of chemistry with sublinear scaling in basis size},
   volume={5},
   ISSN={2056-6387},
   url={http://dx.doi.org/10.1038/s41534-019-0199-y},
   DOI={10.1038/s41534-019-0199-y},
   number={1},
   journal={npj Quantum Information},
   publisher={Springer Science and Business Media LLC},
   author={Babbush, Ryan and Berry, Dominic W. and McClean, Jarrod R. and Neven, Hartmut},
   year={2019},
   month=Nov }

@article{Rubin_2024,
   title={Quantum computation of stopping power for inertial fusion target design},
   volume={121},
   ISSN={1091-6490},
   url={http://dx.doi.org/10.1073/pnas.2317772121},
   DOI={10.1073/pnas.2317772121},
   number={23},
   journal={Proceedings of the National Academy of Sciences},
   publisher={National Academy of Sciences},
   author={Rubin, Nicholas C. and Berry, Dominic W. and Kononov, Alina and Malone, Fionn D. and Khattar, Tanuj and White, Alec and Lee, Joonho and Neven, Hartmut and Babbush, Ryan and Baczewski, Andrew D.},
   year={2024},
   month=May }

@article{Bacon_2006,
   title={Efficient Quantum Circuits for Schur and Clebsch-Gordan Transforms},
   volume={97},
   ISSN={1079-7114},
   url={http://dx.doi.org/10.1103/PhysRevLett.97.170502},
   DOI={10.1103/physrevlett.97.170502},
   number={17},
   journal={Physical Review Letters},
   publisher={American Physical Society (APS)},
   author={Bacon, Dave and Chuang, Isaac L. and Harrow, Aram W.},
   year={2006},
   month=Oct }

@article{li2022circuit,
  title={The circuit design and optimization of quantum multiplier and divider},
  author={Li, Hai-Sheng and Fan, Ping and Xia, Haiying and Long, Gui-Lu},
  journal={Science China Physics, Mechanics \& Astronomy},
  volume={65},
  number={6},
  pages={260311},
  year={2022},
  publisher={Springer}
}

@article{fan2024designs,
  title={Designs of the divider and special multiplier optimizing T and CNOT gates},
  author={Fan, Ping and Li, Hai-Sheng},
  journal={EPJ Quantum Technology},
  volume={11},
  number={1},
  pages={13},
  year={2024},
  publisher={Springer}
}

@article{khattar2025rise,
  title={Rise of conditionally clean ancillae for efficient quantum circuit constructions},
  author={Khattar, Tanuj and Gidney, Craig},
  journal={Quantum},
  volume={9},
  pages={1752},
  year={2025},
  publisher={Verein zur F{\"o}rderung des Open Access Publizierens in den Quantenwissenschaften}
}

@article{gidney2017factoringdirty,
  title={Factoring with n+ 2 clean qubits and n-1 dirty qubits},
  author={Gidney, Craig},
  journal={arXiv preprint arXiv:1706.07884},
  year={2017}
}

@book{brentzimmermann2010modern,
  author    = {Brent, Richard P. and Zimmermann, Paul},
  title     = {Modern Computer Arithmetic},
  publisher = {Cambridge University Press},
  year      = {2010},
  series    = {Cambridge Monographs on Applied and Computational Mathematics},
  volume    = {18},
  url       = {https://maths-people.anu.edu.au/~brent/pub/pub226.html}
}

@inproceedings{cleveFastParallelCircuits2000,
  title={Fast parallel circuits for the quantum Fourier transform},
  author={Cleve, Richard and Watrous, John},
  booktitle={Proceedings 41st annual symposium on foundations of computer science},
  pages={526--536},
  year={2000},
  organization={IEEE}
}

@article{cuccaroNewQuantumRipplecarry2004a,
  title={A new quantum ripple-carry addition circuit},
  author={Cuccaro, Steven A and Draper, Thomas G and Kutin, Samuel A and Moulton, David Petrie},
  journal={arXiv preprint quant-ph/0410184},
  year={2004}
}

@article{kahanamoku-meyerFastQuantumInteger2024,
  title={Fast quantum integer multiplication with zero ancillas},
  author={Kahanamoku-Meyer, Gregory D and Yao, Norman Y},
  journal={arXiv preprint arXiv:2403.18006},
  year={2024}
}

@article{seidelEfficientFloatingPoint2021,
      title={Efficient Floating Point Arithmetic for Quantum Computers}, 
      author={Raphael Seidel and Nikolay Tcholtchev and Sebastian Bock and Colin Kai-Uwe Becker and Manfred Hauswirth},
      journal={arXiv preprint	arXiv:2112.10537},
      year={2021}
}

@inproceedings{haenerQuantumCircuitsFloatingPoint2018,
  title={Quantum circuits for floating-point arithmetic},
  author={Haener, Thomas and Soeken, Mathias and Roetteler, Martin and Svore, Krysta M},
  booktitle={International conference on reversible computation},
  pages={162--174},
  year={2018},
  organization={Springer}
}

@article{mukhopadhyayImprovedQuantumCircuits2026,
  title={Improved quantum circuits for division},
  author={Mukhopadhyay, Priyanka and Gheorghiu, Alexandru and Krovi, Hari},
  journal={arXiv preprint arXiv:2603.18110},
  year={2026}
}

@article{buhrman2025quantum,
  title={Quantum catalytic space},
  author={Buhrman, Harry and Folkertsma, Marten and Mertz, Ian and Speelman, Florian and Strelchuk, Sergii and Subramanian, Sathyawageeswar and Tupker, Quinten},
  journal={arXiv preprint arXiv:2506.16324},
  year={2025}
}

@article{haner2016factoring,
  title={Factoring using 2n+ 2 qubits with Toffoli based modular multiplication},
  author={H{\"a}ner, Thomas and Roetteler, Martin and Svore, Krysta M},
  journal={arXiv preprint arXiv:1611.07995},
  year={2016}
}

@article{berry2019qubitization,
  title={Qubitization of arbitrary basis quantum chemistry leveraging sparsity and low rank factorization},
  author={Berry, Dominic W and Gidney, Craig and Motta, Mario and McClean, Jarrod R and Babbush, Ryan},
  journal={Quantum},
  volume={3},
  pages={208},
  year={2019},
  publisher={Verein zur F{\"o}rderung des Open Access Publizierens in den Quantenwissenschaften}
}

@inproceedings{mogensen2018garbage,
  title = {Garbage-Free Reversible Multiplication and Division},
  booktitle = {Reversible Computation},
  author = {Mogensen, Torben {\AE}gidius},
  editor = {Kari, Jarkko and Ulidowski, Irek},
  year = 2018,
  pages = {253--268},
  publisher = {Springer International Publishing},
  address = {Cham},
  isbn = {978-3-319-99498-7}
}

@misc{gidneyAsymptoticallyEfficientQuantum2019,
      title={Asymptotically Efficient Quantum Karatsuba Multiplication}, 
      author={Craig Gidney},
      year={2019},
      eprint={1904.07356},
      archivePrefix={arXiv},
      primaryClass={quant-ph},
      url={https://arxiv.org/abs/1904.07356}, 
}

@article{regev2009lattices,
  title={On lattices, learning with errors, random linear codes, and cryptography},
  author={Regev, Oded},
  journal={Journal of the ACM (JACM)},
  volume={56},
  number={6},
  pages={1--40},
  year={2009},
  publisher={ACM, New York, NY, USA}
}
\onecolumn

\appendix
\section{Integer multiplication in the phase}\label{sec:appendix:Pgate}
    In this section, we review some of the main results from Kahanamoku-Meyer \& Yao \cite{kahanamoku-meyerFastQuantumInteger2024}. We introduce new notation for expressing these results, notation which we reuse in the subsequent three lemmas.
    
    In Lemma \ref{lem:newPgatedepth}, we show how to perform phase-encoded integer multiplication between two equal-size registers in linear depth. In Lemmas \ref{lem:new P gate complexity two inputs} and \ref{lem:new P gate complexity three inputs}, we show how to perform phase-encoded integer multiplication between differently sized registers, first considering two registers and then three. We give algorithms that are no more expensive than the equal-size case.
    
    To recap, $\p[\alpha]{t}{m}{n}{\phi}$ and $\pover[\alpha]{t}{m}{n}{\phi}$ are defined as follows. Let $x \in \Z_{2^t}, y \in \Z_{2^{m}}, z \in \Z_{2^n}, \phi \in \mathbb{Q}$ and $\alpha \in \{0,1\}$. Then
    \begin{align}
    \p[\alpha]{t}{m}{n}{\phi}:\ket{x}_t\ket{y}_m\ket{z}_{n} &\longmapsto e^{2\pi i \phi yzx^{\alpha}} \ket{x}_t\ket{y}_m\ket{z}_n, \\ \pover[\alpha]{t}{m}{n}{\phi}: \ket{x}_t\ket{y}_m\ket{z}_{n} &\longmapsto e^{2\pi i \phi\rev{y}zx^{\alpha}} \ket{x}_t\ket{y}_m\ket{z}_{n}.
    \end{align}
    It was shown in \cite{kahanamoku-meyerFastQuantumInteger2024} that, for any integer $k\geq2$ of size $O(1)$ with respect to $n$, the operation $\p{0}{n}{n}{\phi}$ requires $O(n^{\log_k(2k-1)})$ gates and no ancilla qubits, and the operation $\p[1]{n}{n}{n}{\phi}$ requires $O(n^{\log_k(3k-2)})$ gates and no ancilla qubits. As a result, for any $\epsilon>0$, we can choose integers $k, k'$ that suppress the asymptotic gate count of both operations to $O(n^{1+\epsilon})$. 

    The algorithms rely on a recursive decomposition of integer multiplication, based on the Toom--Cook method, which we present now.

    \paragraph{Fast integer multiplication.}\label{sec:appendix:Pgate:para:int mult}
    Let $A(\ell, g)$ be a $\ell \times k$ matrix with entries 
    \begin{align}\label{eq:P gate A matrix}
    A(\ell, g)_{i,j} = \begin{cases}
        w_{i}^{k-1-j} \hspace{1em} \text{if $i<g$} \\
        v_{i}^{j} \hspace{3.25em}  \text{if $i\geq g$},
    \end{cases}
    \end{align}
    for integers $w_0, \ldots, w_{g-1}, v_{g}, \ldots, v_{\ell -1}$ of size $O(1)$ with respect to $n$. In other words, \(A(\ell, g)\) is a block matrix formed by vertically stacking two Vandermonde matrices, with the upper block ordered by decreasing powers. No two rows of the matrix are allowed to be the same. Additionally, if $w_{i_1}=-1$ for some $i_1$, $v_{i_2} \neq -1$ for all $i_2$. For the remainder of this recap, we ignore the dependence of this matrix on $g$, since the results hold for any choice of $g$. However, for the subsequent three lemmas, our choice of $g$ becomes relevant.
    
    Consider the function \begin{equation}
    f_{k,d}:\mathbb{Z}_{2^d} \rightarrow \Z_{2^{\lfloor d/k\rfloor + d \bmod k}} \times (\Z_{2^{\lfloor d/k\rfloor}})^{k-1} 
    \end{equation} mapping integers $c \in \Z_{2^d}$ to vectors $(c_{k-1}, \ldots, c_0)$ where $c = \sum_{j=0}^{k-1}(2^{\lfloor d/k\rfloor})^j c_j$. We define the operation \begin{equation}
    \mathcal{A}^{(d)}_{\ell,i}(c) = (A({\ell}) f_{k,d}(c))_i,
    \end{equation} which on input $c \in \Z_{2^d}$ returns the $i^{th}$ entry of the matrix-vector product $A({\ell}) f_{k,d}(c)$. We will often write $\mathcal{A}_{\ell, i}(c)$ when the size of the register the input sits within is clear. Let $\ell_1 = k, \ell_2 = 2k-1, \ell_3 = 3k-2$.   
    It was observed in \cite{kahanamoku-meyerFastQuantumInteger2024} that for $\phi \in \mathbb{Q}$, and $x,y,z \in \Z_{2^n}$,
    \begin{align}
    \begin{split}\label{eq:P gate decomposition of phi}
    \phi z = \sum_{i=0}^{\ell_1 -1} \phi^{(n)}_{\ell_1,i}\mathcal{A}^{(n)}_{\ell_1,i}(z), & \hspace{2em}
    \phi yz = \sum_{i=0}^{\ell_2 -1} \phi^{(n)}_{\ell_2,i} \cdot\mathcal{A}^{(n)}_{\ell_2,i}(y) \cdot \mathcal{A}^{(n)}_{\ell_2,i}(z),\\ 
    \phi xyz = \sum_{i=0}^{\ell_3 -1} &\phi^{(n)}_{\ell_3, i} \cdot \mathcal{A}^{(n)}_{\ell_3,i}(x) \cdot\mathcal{A}^{(n)}_{\ell_3,i}(y) \cdot \mathcal{A}^{(n)}_{\ell_3,i}(z),
    \end{split}
    \end{align}
    where the rational numbers $\phi^{(n)}_{\ell_1, i}, \phi^{(n)}_{\ell_2, i}, \phi^{(n)}_{\ell_3, i}$ depend only on $\phi$, the initial register width $n$, and the matrices $A(\ell_1),$ $A(\ell_2),$ $A(\ell_3)$ respectively. Crucially, this makes them classically pre-computable. For each of the equations, the terms on the right-hand side are of the same form as the initial multiplication on the left-hand side, instead involving integers of size at most $2^{\lfloor n/k\rfloor + O(1)}$. Therefore, we can apply each decomposition recursively. Applying the recursion to depth $O(\log_k n)$ gives integers of size $O(1)$. 

    \paragraph{Recursive algorithms.}\label{sec:appendix:Pgate:para:recursive algorithms}
    These observations reveal recursive algorithms for $\p{0}{n}{n}{\phi}$ and $\p[1]{n}{n}{n}{\phi}$. Suppose for all $i \in \Z_{\ell_2}$, the integers $\mathcal{A}^{(n)}_{\ell_2, i}(y), \mathcal{A}^{(n)}_{\ell_2, i}(z) < 2^{d_2}$ and for all $i\in \Z_{\ell_3}$ the integers $\mathcal{A}^{(n)}_{\ell_3, i }(x), \mathcal{A}^{(n)}_{\ell_3, i}(y), \mathcal{A}^{(n)}_{\ell_3, i}(z) < 2^{d_3}$. Then, to perform $\p{0}{n}{n}{\phi}$ we simply apply $\p{0}{d_2}{d_2}{\phi^{(n)}_{\ell_2, i}}$ to $\ket{\mathcal{A}^{(n)}_{\ell_2,i}(y)}_{d_2}$$\ket{\mathcal{A}^{(n)}_{\ell_2,i}(z)}_{d_2}$ for all $i\in \Z_{\ell_2}$, and to perform $\p[1]{n}{n}{n}{\phi}$ we apply $\p[1]{d_3}{d_3}{d_3}{\phi^{(n)}_{\ell_3, i}}$ to $\ket{\mathcal{A}^{(n)}_{\ell_3,i}(x)}_{d_3}$$\ket{\mathcal{A}^{(n)}_{\ell_3,i}(y)}_{d_3}$ $\ket{\mathcal{A}^{(n)}_{\ell_3,i}(z)}_{d_3}$ for all $i\in \Z_{\ell_3}$. The operations $\p{0}{d_2}{d_2}{\phi_{\ell_2, i}^{(n)}}$ and $\p[1]{d_3}{d_3}{d_3}{\phi_{\ell_3, i}^{(n)}}$ can be implemented in the same way, using $\mathcal{A}_{\ell_2, i}^{(d_2)}(\cdot)$ and $\mathcal{A}_{\ell_3, i }^{(d_3)}(\cdot)$ with newly defined rationals.

    This reduces the task of performing $\p{0}{n}{n}{\phi}$ and $\p[1]{n}{n}{n}{\phi}$ to the task of computing linear combinations of quantum registers, according to the definition of $\mathcal{A}^{(n)}_{\ell, i}(\cdot)$. Since every row of $A(\ell)$ includes at least a single 1, there is always one integer in a linear combination defined by $\mathcal{A}^{(n)}_{\ell,i}(\cdot)$ that does not need to be scaled. Therefore, we can build $\mathcal{A}_{\ell, i}^{(n)}(z)$ on the register containing this integer extended with $O(1)$ clean ancilla qubits, using only out-of-place scaled additions. The ancilla qubit count is clear since we know that $d_2, d_3 =\lfloor n/k\rfloor + O(1)$. An addition scaled by $\lambda$ can be performed with at most $\log(\lambda) +1$ shifted additions. The cost of this sequence of scaled additions on $\ket{z}_n$, for scalars of size $O(1)$ with respect to $n$, is $O(n)$ \cite{cuccaroNewQuantumRipplecarry2004a}. Nothing changes at the lower levels of recursion.
    
    Putting all this together, the algorithm for $\p{0}{n}{n}{\phi}$ starts by partitioning $\ket{y}_n $ and $\ket{z}_n$ into $k$ subregisters according to the definition of $f_{k,n}$. Then it prepares a linear combination of the $y$ subregisters and the $z$ subregisters according to $\mathcal{A}^{(n)}_{\ell_2, i}(\cdot)$, applies $\p{0}{d_2}{d_2}{\phi^{(n)}_{\ell_2, i}}$ to the linear combination pair, and finally reverses the operations returning to the original $\ket{y}_n\ket{z}_n$. It does this sequentially for each $i \in \Z_{\ell_2}$. The algorithm for $\p[1]{n}{n}{n}{\phi}$ is the same, just involving three linear combinations, each sitting on $d_3$ qubits instead. Both algorithms are applied recursively to solve the subproblems, requiring $O(1)$ ancilla qubits at every level of recursion to store each linear combination in a register.
    
    \paragraph{Avoiding ancillas.}\label{sec:appendix:Pgate:para:avoiding ancillas}    
    However, notice that these algorithms do not actually need to store a linear combination of the form ${\mathcal{A}_{\ell,i}^{(n)}(z)}$ in the basis. They just require that the product associated with ${\mathcal{A}_{\ell,i}^{(n)}(z)}$ reaches the phase. Exploiting this fact, it is shown in Appendix C of \cite{kahanamoku-meyerFastQuantumInteger2024} how to get away with only storing $\mathcal{A}^{(n)}_{\ell,i}(z) \bmod 2^{\lfloor n/k \rfloor}$, thereby avoiding the requirement for ancillas at each level of recursion. We review this argument now since we will make a similar one in the proof of Lemma \ref{lem:new P gate complexity three inputs}. 

    Suppose our goal is to push the product $\psi \mathcal{A}_{\ell, i}(y) \mathcal{A}_{\ell, i}(z)$ into the phase of our quantum state, for $\mathcal{A}_{\ell,i}(y), \mathcal{A}_{\ell, i }(z) \in \Z_{2^{d}}$ with $d = \lfloor n/k\rfloor +O(1)$. Instead of applying $\p{0}{d}{d}{\psi}$ to $\ket{\mathcal{A}_{\ell, i}(y)}_d\ket{\mathcal{A}_{\ell, i}(z)}_d$, we can do the following. Consider $(\mathcal{A}_{\ell, i}(y))_{\mathrm{hi}}, (\mathcal{A}_{\ell, i}(z))_{\mathrm{hi}} \in \Z_{2^{d-{\lfloor n/k \rfloor}}}$, and $(\mathcal{A}_{\ell, i}(y))_{\mathrm{lo}}, (\mathcal{A}_{\ell, i}(z))_{\mathrm{lo}} \in \Z_{2^{{\lfloor n/k \rfloor}}}$, such that
    \begin{align}
        \mathcal{A}_{\ell, i}(y) &= 2^{\lfloor n/k \rfloor}(\mathcal{A}_{\ell, i} (y))_{\mathrm{hi}} + (\mathcal{A}_{\ell, i}(y))_{\mathrm{lo}}, \\
        \mathcal{A}_{\ell, i}(z) &= 2^{\lfloor n/k \rfloor}(\mathcal{A}_{\ell, i}(z))_{\mathrm{hi}} + (\mathcal{A}_{\ell, i}(z))_{\mathrm{lo}}.
    \end{align}
    The product can therefore be decomposed as $\psi \mathcal{A}_{\ell, i}(y) \mathcal{A}_{\ell, i}(z)=$
    \begin{equation} \label{eq: Pgate overflow trick}
    =\psi 2^{\lfloor n/k\rfloor}(\mathcal{A}_{\ell, i} (y))_{\mathrm{hi}}\mathcal{A}_{\ell, i}(z) + \psi2^{\lfloor n/k\rfloor}(\mathcal{A}_{\ell, i}(y))_{\mathrm{lo}} (\mathcal{A}_{\ell, i}(z))_{\mathrm{hi}} + \psi(\mathcal{A}_{\ell, i}(y))_{\mathrm{lo}}(\mathcal{A}_{\ell, i}(z))_{\mathrm{lo}}.\end{equation}
    We show how to push each term in Equation \ref{eq: Pgate overflow trick} to the phase, effectively pushing the desired product.

    First, we deal with the term $\psi 2^{\lfloor n/k\rfloor}(\mathcal{A}_{\ell, i} (y))_{\mathrm{hi}}\mathcal{A}_{\ell, i}(z)$. The value $(\mathcal{A}_{\ell, i} (y))_{\mathrm{hi}}$ represents the overflow that occurs when we replace each scaled addition used in the preparation of $\mathcal{A}_{\ell, i} (y)$ with a scaled modular addition, modulo $2^{\lfloor n/k \rfloor}$. We focus on a single scaled addition of the form $y_0 \mapsto y_0 +2^{\lambda_j} y_1$ since each scaled addition of the form $y_0 \mapsto y_0 +\lambda y_1$ can be implemented with at most $\log(\lambda)+1 =O(1)$ additions or subtractions of this form. The overflow contribution of this addition can be placed in two categories: overflow from reducing $2^{\lambda_j} y_1$ modulo $2^{\lfloor n/k \rfloor}$ and overflow from reducing $y_0 + (2^{\lambda_j}  y_1 \bmod 2^{\lfloor n/k \rfloor})$ modulo $2^{\lfloor n/k\rfloor}$. This first type of overflow is determined by the shift $2^{\lambda_j}  $ and the most significant bits of $y_1$. The second type of overflow is the final carry in an addition between two $\lfloor n/k \rfloor$-bit numbers. In the ripple carry addition circuit of \cite{cuccaroNewQuantumRipplecarry2004a}, this final carry is actually stored at the halfway point of the circuit.\footnote{There is an additional subtlety considered in Appendix C of \cite{kahanamoku-meyerFastQuantumInteger2024} that we avoid formally considering here for ease of exposition. The ripple-carry adder of \cite{cuccaroNewQuantumRipplecarry2004a} requires an incoming carry bit. Naively, a clean ancilla qubit is required for this, to ensure that the correct addition is performed. However, it is observed in \cite{kahanamoku-meyerFastQuantumInteger2024} that an idle dirty qubit can be borrowed for the incoming carry instead, and the effect of this qubit on the addition can be pre-emptively uncomputed in the phase using controlled phase rotations. We implicitly apply this technique, which costs no more than the cost to account for overflow, in all of the results that follow.} These facts mean that, using modular additions instead of plain additions does not mean we lose track of the overflow; all of the bits determining this overflow are stored in the basis at some stage regardless, they just may not be stored when the modular addition is complete. Instead of using new ancillas, we can simply control on the relevant bits before (for the first type of overflow) or during (for the second type of overflow) each modular addition, and apply phase rotations to the $z$-register. The angles used for these rotations are determined exclusively by $\mathcal{A}_{\ell, i}(\cdot)$ and $\psi$, thus are classically pre-computable. Since each addition can only have $O(1)$ overflow bits, we only require $O(n)$ controlled phase rotations (for the $n$ qubits of the $z$ register). Letting $y =\sum_{j=0}^{k-1}(2^{\lfloor n/k\rfloor})^j y_j$ and assuming that we build the desired linear combination on $\ket{y_0}_{\lfloor n/k \rfloor }$, this proves that it is possible to map \begin{equation}
    \ket{y}_{n}\ket{z}_n \longmapsto e^{2\pi i \psi2^{\lfloor n/k \rfloor}(\mathcal{A}_{\ell, i} (y))_{\mathrm{hi}}\mathcal{A}_{\ell, i}(z)}\ \ket{y_{k-1}\cdots y_1}_{n- \lfloor n/k \rfloor}\ket{(\mathcal{A}_{\ell, i}(y))_{\rm lo}}_{\lfloor n/k \rfloor}\ket{z}_{n},
    \end{equation}
    using $O(n)$ gates and no ancillas. We note, since it will be relevant in Lemma \ref{lem:new P gate complexity three inputs}, that clearly if $z$ is on a $m$ qubit register, we require $O(m)$ phase rotations to avoid using ancillas and $O(n)$ gates for the sequence of modular additions.
    
    By all of the same reasoning as above, the sequence of additions for $\mathcal{A}_{\ell, i }(z)$ can be replaced with modular additions modulo $2^{\lfloor n/k \rfloor}$ and the overflow $(\mathcal{A}_{\ell ,i }(z))_{\rm hi}$ controlled on. Thus, to push the second term in Equation \ref{eq: Pgate overflow trick} to the phase we require $O(\lfloor n/k\rfloor )$ controlled phase rotations and $O(n)$ gates for the sequence of modular additions. This leaves $\ket{(\mathcal{A}_{\ell, i}(y))_{\mathrm{lo}}}_{\lfloor n/k \rfloor}$ on one of the $k$ subregisters of $\ket{y}_n$ and $\ket{(\mathcal{A}_{\ell, i}(z))_{\mathrm{lo}}}_{\lfloor n/k \rfloor}$ on one of the $k$ subregisters of $\ket{z}_n$. Applying $\p{0}{\lfloor n/k \rfloor}{\lfloor n/k \rfloor}{\psi}$ deals with the third term in Equation \ref{eq: Pgate overflow trick}. This proves that $\p{0}{n}{n}{\phi}$ can be implemented with modular additions and careful accounting of overflow, requiring $O(n)$ gates, and with $\p{0}{\lfloor n/k \rfloor}{\lfloor n/k \rfloor}{\phi_{\ell_2, i}^{(n)}}$ for $i \in \Z_{\ell_2}$. By the master theorem for divide and conquer algorithms, we require $O(n^{\log_k (2k-1)})$ gates and no ancilla qubits.
    
    For $\p[1]{n}{n}{n}{\phi}$, we can avoid ancillas by accounting for the overflow of modular additions using exactly the same techniques. The only difference is that, since the phase-encoded multiplication is between three registers, when we control on $(\mathcal{A}_{\ell, i}(y))_{\rm hi}$, we must apply an operation to $\ket{x}_n\ket{z}_n$ that pushes the product $\psi2^{\lfloor n/k\rfloor}\mathcal{A}_{\ell, i}(x) \mathcal{A}_{\ell, i}(z)$ to the phase instead. The desired operation is simply $k^2 = O(1)$ gates of the form $\p{0}{\lfloor n/k\rfloor} {\lfloor n/k \rfloor}{\psi'}$ where $\psi'$ is determined by $\psi2^{\lfloor n/k \rfloor}$ and $\mathcal{A}_{\ell, i}(\cdot)$. Therefore, the cost of avoiding ancillas at the top-level of recursion is $G(\p{0}{n}{n}{\phi})$. It follows from the master theorem for divide and conquer algorithms that the total recursive algorithm requires  $O(n^{\log_k (3k-2)})$ gates and no ancilla qubits. 

    \paragraph{Lemma conventions.}\label{sec:appendix:Pgate:para:lemma conventions} We briefly mention some notational conventions we will use in the following three lemmas.
    
    It is observed in \cite{kahanamoku-meyerFastQuantumInteger2024} that using the larger register of $\lfloor n/k \rfloor + (n \bmod k)$ qubits for storing linear combinations can be advantageous. Despite this, to simplify our analysis, we will avoid making a register of this form entirely. Instead, at every level of recursion, after we partition a register according to $f_{k,d}$, we will push the product associated with the overhanging $d \bmod k$ qubits into the phase immediately, using the same technique as we used for overflow. By the above analysis, this does not affect the algorithm's asymptotic cost, since $k=O(1)$. This allows us to make the simplifying assumption when performing phase-encoded multiplication between two equal-size registers, that the register size is a power of $k$. 
    
    We drop the modulus when writing a state of the form $\ket{\mathcal{A}_{\ell,i}(z) \bmod 2^{n/k}}_{n/k}$ since it is understood that the integer $\mathcal{A}_{\ell,i}(z)$ must be taken $\bmod \ 2^{n/k}$ to sit on $n/k$ qubits.

    Since the parameters $\phi$ have no impact on the cost of phase-encoded multiplication (in our circuit model; in practice, this is not the case) we omit them from statements involving $G(\cdot)$ and $D(\cdot)$.
    
\newPgatedepth*
\begin{proof}
    The gate and ancilla count claims follow from \cite{kahanamoku-meyerFastQuantumInteger2024}. As described above, the algorithm for $\p{0}{n}{n}{\phi}$ involves sequences of out-of-place modular additions, computing integers of the form $\mathcal{A}^{(n)}_{\ell_2,i}(z) \bmod 2^{n/k }$. For $i \in \Z_{\ell_2}$, we compute (and later uncompute) each integer sequentially. 
    
    Here we will show that it is always possible to parallelise at least two of these computations for $k\geq 3 $. This allows us to perform $\lfloor \ell_2/2 \rfloor$ layers of modular addition sequences at each level of the recursion, instead of $\ell_2$, implying the result for $\p{0}{n}{n}{\phi}$.
    
    Partition the input $\ket{z}_n$ into $k$ subregisters $\ket{z_{k-1}}_{n/k} \cdots \ket{z_0}_{n/k}$ according to $f_{k,n}$ and label the register containing $z_j$ as $Z_j$. In the definition of $A(\ell_2, g)$ from Equation \ref{eq:P gate A matrix}, set $g =\ell_2-1, w_i = i$ for all $i=0, \ldots, \ell_2-2$ and $v_{\ell_2-1} = 0$. Observe that this implies $\mathcal{A}_{\ell_2, 0}(z) = z_0$, $ \mathcal{A}_{\ell_2, \ell_2-1}(z)= z_{k-1}$ and $\mathcal{A}_{\ell_2, 1}(z) = \sum_{j=0}^{k -1}z_j$. Thus, these three linear combinations can be computed simultaneously; simply do nothing to $\ket{z_0}_{n/k}, \ket{z_{k-1}}_{n/k}$ and pick any other register to build $\mathcal{A}_{\ell_2, 1}(z) \bmod 2^{n/k}$ on. Next, consider $\mathcal{A}_{\ell_2, i_1}(z)$ and $\mathcal{A}_{\ell_2, i_2}(z)$ for $i_2 = i_1 +1 <g$. Then
    \begin{equation}
    \mathcal{A}_{\ell_2, i_1}(z) = \sum_{j=0}^{k-1} i_1^{j}z_j, \qquad \mathcal{A}_{\ell_2, i_2}(z) = \sum_{j=0}^{k-1}(i_1+1)^j z_j.
    \end{equation}
    We show how to prepare $\ket{\mathcal{A}_{\ell_2, i_1} (z)}_{n/k}$ and $\ket{\mathcal{A}_{\ell_2, i_2} (z)}_{n/k}$ simultaneously.
    
    Add $i_1Z_{1}$ to $Z_0$. Then add $Z_0$ to $Z_{1}$. This maps \begin{equation} 
    \ket{z_1}_{n/k}\ket{z_0}_{n/k} \longmapsto \ket{(i_1+1)z_1 +z_0}_{n/k}\ket{i_1z_1 + z_0}_{n/k}.
    \end{equation}
    We note that the overflow from $i_1 z_1 + z_0$ must be doubly accounted for at the time of this first addition, since it should be part of the overflow on both $Z_0$ and $Z_1$. Then for $j=2, \ldots, k-1$ add $(i_1)^jZ_j$ to $Z_0$ and add $(i_1 +1)^jZ_j$ to $Z_1$. This prepares $\ket{\mathcal{A}_{\ell_2, i_1} (z)}_{n/k}$ and $\ket{\mathcal{A}_{\ell_2, i_2} (z)}_{n/k}$ on $Z_0$ and $Z_1$ with $2k-2$ out-of-place scaled modular additions. By our choice of $w_2, \ldots, w_{\ell_2-2}$, this means all of the linear combinations $\mathcal{A}_{\ell_2, i}(z)$ for $i=2, \ldots, \ell_2-2$ can be paired up. 

    This gives an algorithm for $\p{0}{n}{n}{\phi}$ that has depth $O(n)$. On input $\ket{y}_n\ket{z}_n$, prepare $\ket{\mathcal{A}_{\ell_2, i}(y)}_{n/k}\ket{\mathcal{A}_{\ell_2, i}(z)}_{n/k}$ for $i=0, 1, \ell_2 -1$, preparing the $y$ register first, then the $z$ register. Then apply the gates $\p{0}{n/k}{n/k}{\phi_{\ell_2, i}^{(n)}}$ to all three pairs of linear combinations simultaneously. Then uncompute the $y$ and $z$ subregisters, and repeat for new indices $i$ which can be paired together as follows: $(2,3), (4,5), \ldots, (\ell_2-3, \ell_2 -2)$ (since $\ell_2$ is odd).

    The new algorithm involves $ (\ell_2-3)/2 +1 = \lfloor \ell_2 /2 \rfloor$ layers of phase-encoded multiplication. Therefore,
    \begin{align}
    D(\mathbf{P}_{0,n,n}) &=  \lfloor(2k-1)/2\rfloor D(\mathbf{P}_{0,n/k,n/k}) +O(n) = O(n).
    \end{align}
\end{proof}

\Pgatecomplexitytwoinputs*
\begin{proof}
    For $m=n$, the result follows from Lemma \ref{lem:newPgatedepth}. Assume that $n = qm +r$ for $q>0$ and $0\leq r <m$. For $r =0$, to perform $\p{0}{m}{n}{\phi}$ on $\ket{y}_{m}\ket{z}_{n}$ we simply partition $\ket{z}_n$ into $q$ subregisters of size $m$, labelled from most to least significant as $Z_{q-1}, \ldots, Z_{0}$, and apply $\p{0}{m}{m}{2^{mi}\phi}$ to $\ket{y}_m$ and $Z_{i}$ for each $i \in \Z_{q}$. This requires $O(q(m)^{1+\epsilon})$ gates, $O(qm)$ depth and no ancilla qubits, implying the result.

    For $r \neq 0$, we instead partition $\ket{z}_n$ into $q$ subregisters of size $m$, and one subregister (corresponding to the most significant qubits) of size $r$. We label these subregisters as before, with the extra subregister labelled $Z_q$, and apply the gates $\p{0}{m}{m}{2^{mi}\phi}$ as before. This reduces the original problem, performing $\p{0}{m}{n}{\phi}$, to the problem of performing $\p{0}{m}{r}{2^{qm}\phi}
    $ on $\ket{y}_m$ and $Z_{q}$. Since we can express $m$ as $m =q'r + r' $ for some $q'>0$ and $0\leq r' <r$, we can apply this construction iteratively. The sequence of register sizes generated follows the Euclidean algorithm for finding the greatest common divisor of \(m\) and \(n\), terminating when one of the divisions has a remainder equal to zero.

    More formally, let $r_0 = n, r_1 =m $ and define $r_{i+1}, r_{i+2}$ for all $i\geq0$ as the integers satisfying $0\leq r_{i+2} < r_{i+1}$ and $r_{i} = q_ir_{i+1} + r_{i+2}$ for some $q_i >0$.
    It follows that the depth and gate count of the described algorithm is
    \begin{align}\label{eq: P gate euclidean depth}
        D(\mathbf{P}_{0,m,n}) &= O(\sum_{i=0}^{\ell} q_ir_{i+1}) = O(n+m -\gcd(n,m)) = O(n), \\ \label{eq: P gate euclidean gate}
        G(\mathbf{P}_{0,m,n}) &= O(\sum_{i=0}^{\ell} q_i(r_{i+1})^{\log_k (2k-1)}) = O(n^{\log_k (2k-1)}),
    \end{align}
    where $r_{\ell+2} =0$. To see Equation \ref{eq: P gate euclidean depth}, observe that $q_ir_{i+1} = r_i - r_{i+2}$, so the entire summation telescopes giving $r_0 +r_1 - r_{\ell+1} -r_{\ell+2}$. Since $r_{\ell+1}$ is by construction the greatest common divisor of $n$ and $m$, the result follows. To see Equation \ref{eq: P gate euclidean gate}, observe that 
    \begin{equation}
        \sum_{i=0}^{\ell} q_i(r_{i+1})^{\log_k (2k-1)} \leq n^{\log_k (2k-1)-1}\sum_{i=0}^{\ell} q_i(r_{i+1}) = O(n^{\log_k (2k-1)}).
    \end{equation}
    Thus, for any $\epsilon >0$, $\p{0}{m}{n}{\phi}$ can be performed using $O(n^{1+\epsilon})$ gates, $O(n)$ depth and no ancilla qubits.
\end{proof}
\Pgatecomplexitythreeinputs*
\begin{proof}
    First, we note that proving the claim for $\p[\alpha]{t}{m}{n}{\phi}$ immediately implies that it holds for $\pover[\alpha]{t}{m}{n}{\phi}$; for every operation involving the $j^{th}$ qubit of $\ket{y}_m$ in the quantum circuit for $\p[\alpha]{t}{m}{n}{\phi}$, perform that operation with the $(m-j-1)^{th}$ qubit of $\ket{y}_m$ instead.
    
    For $t=0$ and $\alpha=0$, the gate count, depth and ancilla count claims follow from Lemma \ref{lem:new P gate complexity two inputs}. Assume that $t \neq 0$ and $\alpha =1$. We prove the result by performing three separate recursions on the input $\ket{x}_t\ket{y}_m\ket{z}_n$, using Equations \ref{eq:P gate decomposition of phi}.

    Our goal is to push the product $\phi x y z$ into the phase of our input state. For now, assume that $n$ is a power of $k$. The first step of the algorithm reduces the problem of performing a phase multiplication between three registers of sizes $t,m$ and $n$, to the problem of performing a phase multiplication between two equal-size registers, that are close in size to $m$, and a register of size $t$. The second step of the algorithm reduces this further to the problem of performing a multiplication between three equal-size registers, all of which are close in size to $t$. The third step applies the phase multiplication algorithm of \cite{kahanamoku-meyerFastQuantumInteger2024} to these three equal-size registers. We make this argument more precise below.
    
    Recall that $\ell_1 = k, \ell_2 = 2k-1$ and $\ell_3 = 3k-2$. We use the compact notation
    \begin{equation}
    [\mathcal{A}]_{\ell, i}^{(n/k^v)} (z) = \mathcal{A}_{\ell, i}^{(n/k^v)}(z) \bmod 2^{n/k^{v+1}},
    \end{equation} for a modulo-reduced linear combination. We start by partitioning $\ket{z}_n$ into $k$ subregisters according to $f_{k,n}$ and preparing modulo-reduced linear combinations of these subregisters sequentially. We follow the decomposition
    \begin{equation}
    \phi z = \sum_{i=0}^{\ell_1 -1} \phi^{(n)}_{\ell_1,i} \mathcal{A}^{(n)}_{\ell_1,i}(z)
    \end{equation}
    from Equations \ref{eq:P gate decomposition of phi}, accounting for overflow in the phase directly as we are used to. The cost of accounting for the overflow is $G(\mathbf{P}_{0,t,m})$ and the cost to prepare the linear combinations is $O(n)$. 

    This reduces the problem of pushing $\phi xyz$ to the phase, to the problem of pushing
    \begin{equation}\label{eq: new P gate our decomp 1}
        \sum_{i \in \Z_{\ell_1}} \phi_{\ell_1, i}^{(n)} \cdot [\mathcal{A}]_{\ell_1, i}^{(n)} (z)  \cdot  xy
    \end{equation}
    to the phase instead, which can be solved by applying the gates $\p[1]{t}{m}{n/k}{\phi^{(n)}_{\ell_1,i}}$ for $i \in \Z_{\ell_1}$ exactly when each modulo-reduced linear combination is being stored in a register. 
    
    Each subproblem can be reduced to $\ell_1$ new subproblems involving smaller phase multiplications, via $\mathcal{A}_{\ell_1, i }^{(n/k)}$. We can repeat this process until we have $\ell_1^{s}$ subproblems for $s = \lfloor \log_k (n/m)\rfloor$. We describe this new problem more formally next.
    
    Let $\mathbf{i} = (i_0, \ldots, i_{s-1}) \in \Z_{\ell_1}^{s}$. We define 
    \begin{equation}
        z_{\mathbf{i}} = [\mathcal{A}]_{\ell_1, i_{s-1}}^{(n/k^{s-1})} \circ [\mathcal{A}]_{\ell_1, i_{s-2}}^{(n/k^{s-2})} \circ \cdots \circ [\mathcal{A}]_{\ell_1, i_1}^{(n/k)} \circ [\mathcal{A}]_{\ell_1, i_0}^{(n)} (z).
    \end{equation}
    This integer is derived by following the recursion branch $\mathbf{i}$. It sits on $n/k^s$ qubits, one of the now $k^s$ subregisters of the original register $\ket{z}_n$. Let $\phi_{\mathbf{i}}$ denote the corresponding pre-computable rational, determined by the branch, the original register size $n$, the rational $\phi$, and the matrix $A(\ell_1)$.
    We have reduced the problem of pushing $\phi xyz$ to the phase, to the problem of pushing
    \begin{equation}
        \sum_{\mathbf{i} \in \Z_{\ell_1}^s}\phi_{\mathbf{i}}z_{\mathbf{i}} xy.
    \end{equation}
    to the phase instead. This problem is solved by applying $\ell_1^s$ gates $\p[1]{t}{m}{n/k^s}{\phi_{\mathbf{i}}}$ to $\ket{x}_t\ket{y}_m\ket{z_{\mathbf{i}}}_{n/k^s}$.

    By the definition of $s$, $km > n/{k^s} \geq m$. At this point in the algorithm, we would like to apply the decomposition of $\phi yz$ given in Equations \ref{eq:P gate decomposition of phi} to $\phi_{\mathbf{i}}z_{\mathbf{i}}y$. The only trouble is that this decomposition, requires that $z_{\mathbf{i}}$ and $y$ sit on registers of the same size. To get around this fact, we imagine that the bitstring for $y$ is padded with $n/k^s -m$ zeroes, allowing us to apply $\mathcal{A}_{\ell_2, j}^{(n/k^s)}(\cdot)$ to this new bitstring. It follows that
    \begin{equation}
        \phi_{\mathbf{i}}z_{\mathbf{i}} y
    =    \sum_{j \in \Z_{\ell_2}}(\phi_{\mathbf{i}})_{\ell_2, j}^{(n/k^s)} \cdot \mathcal{A}_{\ell_2, j}^{(n/k^s)}(z_{\mathbf{i}}) \cdot \mathcal{A}_{\ell_2, j}^{(n/k^s)}(y).
    \end{equation}

    The challenge, as usual, becomes to reduce this to a problem involving linear combinations of $z_\mathbf{i}$ and $y$ that sit on $n/k^{s+1}$ qubits, accounting for overflow directly. There are two subtleties that make this different from the equal-size register case discussed in the paragraph on \nameref{sec:appendix:Pgate:para:avoiding ancillas}.

    First, we cannot build the linear combination $\mathcal{A}_{\ell_{2}, j}^{(n/k^{s})}(y)$ on any subregister of the $m$ qubit register $\ket{y}_m$, since $m$ is not being partitioned into $k$ equal size subregisters, but into registers of size $n/k^{s+1}$ and one register of size $m \bmod n/k^{s+1}$. The only guarantee we have is that there exists at least one subregister of size $n/k^{s+1}$, the least significant subregister. Therefore, we must build the linear combination on this subregister. Choosing $g=\ell_2$ in the definition of $A(\ell_2, g)$, means that this is always possible using just out-of-place scaled additions.

    Second, as we just noted, one of the subregisters of $\ket{y}_{m}$ will have size $m \bmod n/k^{s+1}$. Let $d = m \bmod n/k^{s+1}$ This means $\mathcal{A}_{\ell_2, j}^{(n/k^{s})}(y)$ involves a scaled addition of the form $y_0 \mapsto y_0 + \lambda y_h$ for $y_0$ sitting on $n/k^{s+1}$ qubits and $y_h$ sitting on $d$ qubits. We propose to first perform all other scaled modular additions between subregisters of $\ket{y}_m$, so that on the least significant $n/k^{s+1}$ qubits of $\ket{y}_m$ we have $[\mathcal{A}]_{\ell_2, j}^{(n/k^{s})}(y) - \lambda y_h \bmod 2^{n/k^{s+1}}$. 
    We use the compact notation
    \begin{equation}\label{eq: new P gate mod lin comb minus}
        [\mathcal{A}]_{\ell_2, j}^{(n/k^s)}(y, - y_h) = [\mathcal{A}]_{\ell_2, j}^{(n/k^{s})}(y) - \lambda y_h \bmod 2^{n/k^{s+1}}
    \end{equation}
    to mean the modulo-reduced linear combination $[\mathcal{A}]_{\ell_2, j}^{(n/k^s)}(y)$ minus the term in this linear combination that involves the integer $y_h$, with everything taken modulo $2^{n/k^{s+1}}$. 
    
    We then put $\lambda y_h - (\lambda y_h \bmod 2^d)$ into the phase in the usual way requiring $G(\mathbf{P}_{0,t,n/k^s})$ gates, and we add $\lambda y_h \bmod 2^{d}$ to the least significant $d$ qubits of the $n/k^{s+1}$ qubit subregister containing $[\mathcal{A}]_{\ell_2, j}^{(n/k^s)}(y, - y_h)$, using the modular addition circuit of \cite{cuccaroNewQuantumRipplecarry2004a}. We pause the circuit midway to control on the overflow, which we denote $c_{y_h}$, of this addition, at a cost of $G(\mathbf{P}_{0,t,n/k^{s}})$. Since the overflow we are accounting for is not $\bmod \ 2^{n/k^{s+1}}$ rather it is $\bmod \ 2^{d}$, the $n/k^{s+1}$ qubit subregister that we have been building on now contains $[\mathcal{A}]_{\ell_2, j}^{(n/k^{s})}(y, -y_h) +(\lambda y_h \bmod 2^d) - 2^{d}c_{y_h}.$

    Since all overflow can consistently be accounted for directly in the phase, we have successfully reduced the problem of pushing $\phi_{\mathbf{i}}z_{\mathbf{i}}y$ to the phase to the problem of pushing 
    \begin{equation}\label{eq: new P gate our decomp 2}
        \sum_{j \in \Z_{\ell_2}}(\phi_{\mathbf{i}})_{\ell_2, j}^{(n/k^s)} \cdot [\mathcal{A}]_{\ell_2, j}^{(n/k^s)}(z_{\mathbf{i}}) \cdot \left([\mathcal{A}]_{\ell_2, j}^{(n/k^s)}(y, -y_h) + (\lambda y_h \bmod 2^d) - 2^dc_{y_h}\right)
    \end{equation}
    to the phase, which is possible at this stage in the algorithm with $\ell_1^{s}\ell_2$ gates of the form $\p[1]{t}{n/k^{s+1}}{n/k^{s+1}}{(\phi_{\mathbf{i}})_{\ell_2, j}^{(n/k^s)}}$.
    
    Each term in Equation \ref{eq: new P gate our decomp 2} can be decomposed using $\mathcal{A}_{\ell_2, j}^{(n/k^{s+1})}(\cdot)$, preparing $\ell_1^{s}\ell_2^2$ subproblems, and so on, until we have $\ell_1^{s}\ell_2^r$ subproblems for $r = \lfloor \log_k (n/tk^s)\rfloor$. We formalise this argument next.

    Let $\mathbf{j} = (j_0, \ldots, j_{r-1}) \in \Z_{\ell_2}^{r}$. We define 
    \begin{equation}
        z_{\mathbf{i},\mathbf{j}} = [\mathcal{A}]_{\ell_2, j_{r-1}}^{(n/k^{s+r-1})} \circ \cdots \circ [\mathcal{A}]_{\ell_2, j_0}^{(n/k^s)} (z_{\mathbf{i}})
    \end{equation}
    and 
    \begin{equation}
        y_{\mathbf{j}} = [\mathcal{A}]_{\ell_2, j_{r-1}}^{(n/k^{s+r-1})} \circ \cdots \circ [\mathcal{A}]_{\ell_2, j_{1}}^{(n/k^{s+1})} \left([\mathcal{A}]_{\ell_2, j_{0}}^{(n/k^s)} (y, -y_h) + (\lambda y_h \bmod 2^d) - 2^d c_{y_h}\right).
    \end{equation}
    Both integers sit on $n/k^{s+r}$ qubits. $z_{\mathbf{i}, \mathbf{j}}$ corresponds to the branch $\mathbf{i}$ from the first recursion and the branch $\mathbf{j}$ from the second recursion while $y_{\mathbf{j}}$ corresponds to the branch $\mathbf{j}$ from the second recursion. Let $\phi_{\mathbf{i}, \mathbf{j}}$ be the corresponding pre-computable rational, determined by the branches $\mathbf{i}$ and $\mathbf{j}$, the original register size $n$, the rational $\phi$, and the matrices $A(\ell_1)$ and $A(\ell_2)$.

    We have reduced the problem of pushing $\phi xyz$ to the phase, to the problem of pushing
    \begin{equation}
        \sum_{\mathbf{i} \in \Z_{\ell_1}^s, \mathbf{j} \in \Z_{\ell_2}^r}\phi_{\mathbf{i}, \mathbf{j}} z_{\mathbf{i}, \mathbf{j}} y_\mathbf{j} x
    \end{equation}
    to the phase. This problem can be solved by applying $\ell_1^{s}\ell_2^r$ gates $\p[1]{t}{n/k^{s+r}}{n/k^{s+r}}{\phi_{\mathbf{i}, \mathbf{j}}}$. Again at this point in the algorithm, we would like to apply the decomposition of $\phi xyz$ given in Equations \ref{eq:P gate decomposition of phi}, but we cannot directly, since $x$ sits on $t$ qubits, whereas $z_{\mathbf i, \mathbf{j}}$ and $y_{\mathbf{j}}$ both sit on $n/k^{s+r}$ qubits. 
    
    Since $kt > n/k^{s+r} \geq t$, we know that we can consider the padded version of the bitstring $x$ with $n/k^{s+r} -t$ zeroes and apply $\mathcal{A}_{\ell_3, g}^{(n/k^{s+r})}(\cdot)$ to this padded bitstring. For all the same reasons as we gave when applying $\mathcal{A}_{\ell_{2}, j}^{(n/k^s)}(\cdot)$ to the padded version of $y$, the problem of pushing $\phi_{\mathbf{i}, \mathbf{j}} z_{\mathbf{i}, \mathbf{j}} y_\mathbf{j} x$ to the phase can be reduced to the problem of pushing 
    \begin{equation}
    \begin{split}
        &\sum_{g \in \Z_{\ell_3}}(\phi_{\mathbf{i}, \mathbf{j}})_{\ell_3, g}^{(n/k^{s+r})} \cdot [\mathcal{A}]_{\ell_3, g}^{(n/k^{s+r})} (z_{\mathbf{i}, \mathbf{j}}) \cdot [\mathcal{A}]_{\ell_3, g}^{(n/k^{s+r})} (y_\mathbf{j}) \cdot  \\ &\hspace{5em}\cdot\left([\mathcal{A}]_{\ell_3, g}^{(n/k^{s+r})} (x, -x_h) + (\lambda x_h \bmod 2^d) -2^d c_{x_h}\right)
    \end{split}
    \end{equation}
    to the phase, where here $d = t \bmod n/k^{s+r+1}$, $[\mathcal{A}]_{\ell_3, g}^{(n/k^{s+r})} (x, -x_h)$ is defined by Equation \ref{eq: new P gate mod lin comb minus}, $x_h$ is the integer contained in the register of $d$ qubits, $\lambda$ is the scalar required for the addition involving $x_h$, and $c_{x_h}$ is the overflow from adding $\lambda x_h \bmod 2^d$ to $[\mathcal{A}]_{\ell_3, g}^{(n/k^{s+r})} (x, -x_h) \bmod 2^d$. This problem can be solved with $\ell_1^{s}\ell_2^{r}\ell_3$ gates of the form \begin{equation}
        \p[1]{n/k^{s+r+1}}{n/k^{s+r+1}}{n/k^{s+r+1}}{(\phi_{\mathbf{i}, \mathbf{j}})_{\ell_3, g}^{(n/k^{s+r})}}.
    \end{equation} 
    Each phase multiplication is between three equal-size registers and can be performed using the method of \cite{kahanamoku-meyerFastQuantumInteger2024}.

    If $n$ is not a power of $k$ (recall that we assumed it was for simplicity at the start), at the top level of recursion we can use $k$ subregisters of size $\lfloor n/k \rfloor$ instead and push the product that involves the most significant $n \bmod k$ qubits to the phase directly, as discussed in \nameref{sec:appendix:Pgate:para:lemma conventions}. We can repeat this at every level of the recursion. This is no more expensive than the cost of accounting for overflow at each level. 
    Thus, whether $n$ is a power of $k$ or not does not affect the asymptotic cost of the algorithm.
    
    Finally, we analyse this cost. We plug back in the values $\ell_1 = k, \ell_2 = 2k-1, \ell_3 = 3k-2$, $s = \lfloor \log_k (n/m)\rfloor$, and $r = \lfloor \log_k(n/tk^s)\rfloor$. The recurrence relation determining the cost of the first recursion is
    \begin{equation}
        G(\mathbf{P}_{t,m,n}) = kG(\mathbf{P}_{t,m,n/k}) + O(n) + G(\mathbf{P}_{0,t,m}).
    \end{equation}
    Recursing until we have $\ell_1^s$ subproblems means the gate count of this first recursion is \begin{equation}
        O(n s) + O(k^s m^{\log_{k}(2k-1)}) = O(n \log(n/m)) + O((n/m) m^{\log_{k}(2k-1)})
    \end{equation} using Lemma \ref{lem:new P gate complexity two inputs}.
    We then showed that
    \begin{equation}
        G(\mathbf{P}_{t,m,n/k^s}) = (2k-1)G(\mathbf{P}_{t,n/k^{s+1},n/k^{s+1}}) + O(n/k^{s}) + G(\mathbf{P}_{0,t,n/k^{s}})
    \end{equation}
    Recursing until we have $\ell_1^{s}\ell_2^{r}$ subproblems means the gate count of the second recursion is \begin{equation}
    O(k^s(n/k^s)^{\log_k (2k-1)} r) = O((n/m)m^{\log_k (2k-1)}\log (m/t)).
    \end{equation}
    Finally, we showed that $G(\mathbf{P}_{t,n/k^{s+r},n/k^{s+r}}) =$
    \begin{equation}
         = (3k-2)G(\mathbf{P}_{n/k^{s+r+1},n/k^{s+r+1},n/k^{s+r+1}}) + O(n/k^{s+r}) + G(\mathbf{P}_{0,n/k^{s+r},n/k^{s+r}})
    \end{equation}
    Recursing until we have $\ell_1^{s}\ell_2^r\ell_3^q$ subproblems where $q = \lfloor \log_k (n/k^{s+r}) \rfloor$ means that the gate count of this third recursion is 
    \begin{equation}
    O(k^{s}(2k-1)^{r} (n/k^{s+r})^{\log_k (3k-2)}) =
    O((n/m)(m/t)^{\log_k (2k-1)} t^{\log_k (3k-2)}).
    \end{equation}
    Putting all of these recursions together, it is clear that $\p[1]{t}{m}{n}{\phi}$ can be implemented using $O(n^{\log_k (3k-2)})$ gates. The algorithm also uses no ancillas.
    It follows that for any $\epsilon >0$, we can pick an integer $k$ such that $\p[\alpha]{t}{m}{n}{\phi}$ can be performed using $O(n^{1 +\epsilon})$ gates and no ancilla qubits.
\end{proof}

\begin{remark}
    Suppressing the gate count of these algorithms to get arbitrarily close to linear comes at the cost of large, impractical constant factors. It was estimated in \cite{kahanamoku-meyerFastQuantumInteger2024} that, for multiplication at cryptographically relevant integer sizes, choices of $k$ leading to $O(n^{1.30\ldots})$ gates for $\p{0}{n}{n}{\phi}$ and $O(n^{1.46\ldots})$ gates for $\p[1]{n}{n}{n}{\phi}$ would be practical. Determining choices of $k$ that lead to practical gate counts for $\p{0}{m}{n}{\phi}$ and $\p[1]{t}{m}{n}{\phi}$ requires separate estimates.
\end{remark}

\section{The controlled and parametrised quantum Fourier transform}\label{sec:appendix:qft}
\qftgatecomplexity*
\begin{proof}
First, we prove the claim for $m=0$. To recap, $\qft{0}{n}{\phi}$ is defined as follows. Let $ y\in \Z_{2^n}$ and $\phi \in \Z_{2^n}^{\times}$. Then,
\begin{align}
\qft{0}{n}{\phi}: \ket{y}_n \longmapsto \frac{1}{\sqrt {2^{n}}} \sum_{z \in \Z_{2^n}}e^{2\pi i \phi yz/2^n}\ket{\rev{z}}_n.
\end{align}
We prove that $\qft{0}{n}{\phi}$ can be defined recursively as
\[
\qft{0}{n}{\phi} =  \left(\mathbf{I}_{\lceil n/2 \rceil } \otimes \qft{0}{\lfloor n/2 \rfloor}{\phi}\right) \circ \left(\pover{0}{\lceil n/2 \rceil}{\lfloor n/2 \rfloor}{\phi/2^{n}}\right)\circ \left(\qft{0}{\lceil n/2 \rceil}{\phi} \otimes \mathbf{I}_{\lfloor n/2 \rfloor}\right).
\]

For $y \in \Z_{2^n}$, consider $y_1 \in \Z_{2^{\lceil n/2 \rceil }}, y_2 \in \Z_{2^{\lfloor n/2 \rfloor}}$ such that $y = 2^{\lfloor n/2 \rfloor}y_1 + y_2$. Then consider the following algorithm on input $\ket{y}_n= \ket{y_1}_{\lceil n/2 \rceil}\ket{y_2}_{\lfloor n/2\rfloor}$,
\begin{flalign}
&\xrightarrow{\qft{0}{\lceil n/2 \rceil}{\phi} \otimes \mathbf{I}_{\lfloor n/2 \rfloor}} 
\frac{1}{\sqrt{2^{\lceil n/2 \rceil }}} \sum_{\substack{z_2 \in \Z_{2^{\lceil n/2 \rceil }}}} e^{2\pi i \phi( 2^{\lfloor n/2 \rfloor}y_1 z_2) /2^n} \ket{\rev{z_2}}_{\lceil n/2 \rceil}\ket{y_2}_{\lfloor n/2\rfloor} &&
\end{flalign}
\begin{flalign}
&\xrightarrow{\pover{0}{\lceil n/2 \rceil}{\lfloor n/2 \rfloor}{\phi/2^{n}}} 
\frac{1}{\sqrt{2^{\lceil n/2 \rceil }}} \sum_{\substack{z_2 \in \Z_{2^{\lceil n/2 \rceil }}}} e^{2\pi i \phi( 2^{\lfloor n/2 \rfloor}y_1 z_2 + y_2z_2) /2^n} \ket{\rev{z_2}}_{\lceil n/2 \rceil}\ket{y_2}_{\lfloor n/2\rfloor} &&
\end{flalign}
\begin{flalign}
&\xrightarrow{\mathbf{I}_{\lceil n/2 \rceil } \otimes \qft{0}{\lfloor n/2 \rfloor}{\phi}} \frac{1}{\sqrt{2^n}} \sum_{\substack{z_1 \in \Z_{2^{\lfloor n/2 \rfloor }}, \\z_2 \in \Z_{2^{\lceil n/2 \rceil }}}} e^{2\pi i \phi( 2^{\lfloor n/2 \rfloor}y_1 z_2+ 2^{\lceil n/2 \rceil}z_1y_2 + y_2z_2) /2^n} \ket{\rev{z_2}}_{\lceil n/2 \rceil}\ket{\rev{z_1}}_{\lfloor n/2\rfloor}&& 
\end{flalign}
\begin{flalign}
&= \frac{1}{\sqrt{2^n}} \sum_{\substack{z_1 \in \Z_{2^{\lfloor n/2 \rfloor }}, \\z_2 \in \Z_{2^{\lceil n/2 \rceil }}}} e^{2\pi i \phi( 2^{\lfloor n/2 \rfloor}y_1 + y_2)(2^{\lceil n/2 \rceil}z_1 + z_2) /2^n} \ket{\rev{z_2}}_{\lceil n/2 \rceil}\ket{\rev{z_1}}_{\lfloor n/2\rfloor}. &&
\end{flalign}
Since the map $(\rev{z_2}, \rev{z_1}) \mapsto 2^{\lfloor n/2 \rfloor}\rev{z_2} + \rev{z_1} = \rev{2^{\lceil n/2 \rceil}z_1 + z_2}$ is a one-to-one correspondence between $\Z_{2^{\lceil n/2 \rceil }} \times \Z_{2^{\lfloor n/2 \rfloor}}$ and $\Z_{2^n}$, we can rewrite this as
\begin{flalign}
&= \frac{1}{\sqrt{2^n}} \sum_{z \in \Z_{2^n}} e^{2\pi i \phi yz/2^n} \ket{\rev{z}}_n. &&
\end{flalign}
Therefore, the problem of implementing $\qft{0}{n}{\phi}$ reduces to the problem of implementing the base case of this recursion $\qft{0}{1}{\phi}$ and $\pover{0}{\lceil t/2 \rceil}{\lfloor t/2 \rfloor}{\phi/2^t}$ for any $t$ in the range $2\leq t\leq n$.  

Since $\phi \in \Z_{2^n}^{\times}$, the least significant bit of $\phi$ equals 1. Thus, \begin{equation}
\qft{0}{1}{\phi} \ket{y_k}_1 = \frac{1}{\sqrt2}\left(\ket{0} + e^{2\pi i \phi y_k/2} \ket{1}\right)= \frac{1}{\sqrt2}\left(\ket{0} + e^{2\pi i y_k/2} \ket{1}\right) = H\ket{y_k}_1.
\end{equation} So the base case $\qft{0}{1}{\phi}$ is just a Hadamard gate. From Lemma \ref{lem:new P gate complexity two inputs} we know how to implement each of the gates $\pover{0}{\lceil t/2 \rceil}{\lfloor t/2 \rfloor}{\phi/2^t}$ for $2\leq t\leq n$.
Thus, by the recursive definition above and Lemma \ref{lem:new P gate complexity two inputs}, it follows that, for any $\epsilon >0$,
\begin{align}G(\qft{0}{n}{\phi}) &\leq  2 G(\qft{0}{\lceil n/2 \rceil}{\phi}) + O(\lceil n/2\rceil^{1+\epsilon}) = O(n^{1+\epsilon}) \\ \intertext{and}
D(\qft{0}{n}{\phi}) &\leq  2 D(\qft{0}{\lceil n/2 \rceil}{\phi}) + O(\lceil n/2 \rceil) = O(n\log n).
\end{align}

Next, we prove the claim for arbitrary $m$. Using the recursive definition for $\qft{0}{n}{\phi}$, we give a recursive definition for $\qft[\alpha]{m}{n}{\phi}$. To recap, $\qft[\alpha]{m}{n}{\phi}$ is defined as follows.
Let $x \in \Z_{2^m}^{\times}, y \in \Z_{2^n}$ and $\phi \in \Z_{2^n}^{\times}$. Let $\alpha \in \{0,1\}$. Then
\begin{align}
\qft[\alpha]{m}{n}{\phi}: \ket{x}_m\ket{y}_n \longmapsto \ket{x}_m\frac{1}{\sqrt {2^{n}}} \sum_{z \in \Z_{2^n}}e^{2\pi i \phi yzx^{\alpha}/2^n}\ket{\rev{z}}_n.
\end{align}
$\qft[0]{m}{n}{\phi}$ requires the same gate count and depth as $\qft{0}{n}{\phi}$ since $\qft[0]{m}{n}{\phi} = \mathbf{I}_{m} \otimes \qft{0}{n}{\phi}$. As for $\qft[1]{m}{n}{\phi}$, controlling on the recursive definition of $\qft{0}{n}{\phi}$ gives $\qft[1]{m}{n}{\phi}=$
\[ = \left(\mathbf{I}_{\lceil n/2 \rceil } \otimes \qft[1]{m}{\lfloor n/2 \rfloor}{\phi}\right) \circ \left(\pover[1]{m}{\lceil n/2 \rceil}{\lfloor n/2 \rfloor}{\phi/2^{n}}\right)\circ \left(\qft[1]{m}{\lceil n/2 \rceil}{\phi} \otimes \mathbf{I}_{\lfloor n/2 \rfloor}\right).
\]
Since $e^{2\pi i \psi x/2^{t}} = e^{2\pi i \psi (x \bmod 2^t)/2^t}$ for any integer $\psi$, it is clear that controlling on all $m$ qubits of $\ket{x}_m$ in the above operations is not always necessary. If $m\geq \lfloor n/2\rfloor$ or $m\geq \lceil n/2\rceil$, then we just need to control on the least significant $\lfloor n/2 \rfloor$ or $\lceil n/2 \rceil$ qubits when performing $\qft[1]{m}{\lfloor n/2\rfloor}{\phi}$ or $\qft[1]{m}{\lceil n/2 \rceil}{\phi}$. Similarly if $m \geq n$, we should control on just the $n$ least significant qubits of $\ket{x}_m$ when performing $\pover[1]{m}{\lceil n/2 \rceil}{\lfloor n/2 \rfloor}{\phi/2^n}$. It follows that,
\begin{align*}
    &\qft[1]{m}{n}{\phi} = \left(\mathbf{I}_{\lceil n/2 \rceil } \otimes \qft[1]{\min\{m, \lfloor n/2 \rfloor\}}{\lfloor n/2 \rfloor}{\phi}\right) \circ \\
    &\hspace{4em} \circ \left(\pover[1]{\min\{m, n\}}{\lceil n/2 \rceil}{\lfloor n/2 \rfloor}{\phi/2^{n}}\right)\circ \left(\qft[1]{\min\{m, \lceil n/2 \rceil\}}{\lceil n/2 \rceil}{\phi} \otimes \mathbf{I}_{\lfloor n/2 \rfloor}\right).
\end{align*}
Therefore, the problem of implementing $\qft[1]{m}{n}{\phi}$ reduces to the problem of implementing the base case of this recursion $\qft[1]{1}{1}{\phi}$
and $\pover{s}{\lceil t/2 \rceil}{\lfloor t/2 \rfloor}{\phi/2^t}$ for any $t$ in the range $2\leq t \leq n$ and any $s$ in the range $1\leq s\leq t$.

Since $x \in \Z_{2^m}^{\times}$ and $\phi \in \Z_{2^n}^{\times}$, $\qft[1]{1}{1}{\phi}\ket{x_0}_1\ket{y_k}_1 =$
\begin{equation}
= \ket{x_0}_1\frac{1}{\sqrt2}\left(\ket{0} + e^{2\pi i \phi x_0y_k/2} \ket{1}\right)= \ket{x_0}_1\frac{1}{\sqrt2}\left(\ket{0} + e^{2\pi i y_k/2} \ket{1}\right) = \ket{x_0}_1H\ket{y_k}_1.
\end{equation} So the base case $\qft[1]{1}{1}{\phi}$ is again just a Hadamard gate. From Lemma \ref{lem:new P gate complexity three inputs} we know how to implement each of the gates $\pover{s}{\lceil t/2 \rceil}{\lfloor t/2 \rfloor}{\phi/2^t}$ for $2\leq t \leq n$ and $1\leq s \leq t$.
Thus, by the recursive definition above and Lemma \ref{lem:new P gate complexity three inputs}, it follows that, for any $\epsilon >0$,
\begin{align}G(\qft[1]{m}{n}{\phi}) &\leq  2 G(\qft[1]{\lceil n/2 \rceil}{\lceil n/2 \rceil}{\phi}) + O( n^{1+\epsilon}) = O(n^{1+\epsilon}).
\end{align}
\end{proof}
\begin{remark}
    For the same choice of parameter $k$ (using the notation of Appendix \ref{sec:appendix:Pgate}), $\qft{0}{n}{\phi}$ and $\qft[1]{m}{n}{\phi}$ have different asymptotic gate counts just like $\p{0}{m}{n}{\phi}$ and $\p[1]{t}{m}{n}{\phi}$. However, by the above result, for any $\epsilon>0$, we can pick an integer $k$ that ensures $G(\qft[\alpha]{m}{n}{\phi}) = O(n^{1+\epsilon})$.
\end{remark}

\section{Computing the trailing zero count coherently}\label{sec:appendix:additional}
\trailingzerocount*
\begin{proof}
    To recap, $\mathbf{TZ}_{m,\lceil\log (m+1)\rceil}$ is defined as follows. For $x\in \Z_{2^m}$, $\text{tz}(x)$$ =\log (\gcd(x, 2^{m}))$,
    \begin{align}
        \mathbf{TZ}_{m,\lceil\log (m+1)\rceil}:\ket{x}_{m}\ket{0}_{\lceil\log (m+1)\rceil} \longmapsto \ket{x}_{m}\ket{\text{tz}(x)}_{\lceil\log (m+1)\rceil}.
    \end{align}
    This operation counts the number of trailing zeroes $\text{tz}(x)$ of the bitstring $x$.

    Label the registers $\ket{x}_m$ and $\ket{0}_{\lceil\log (m+1)\rceil}$, as $X$ and $Y$.
    Let $\mathrm{CX}(X_i \to Y_j)$ denote a CNOT gate that is controlled on $i^{th}$ qubit of $X$ with the $j^{th}$ qubit of $Y$ as the target. Let $\mathrm{CCX}(X_i, X_k \to Y_j)$ denote a Toffoli gate with controls $X_i, X_k$ and target $Y_j$. We use $\widehat{X_i}$ to indicate that a gate controlled on $X_i$ is negatively controlled.

    We prove the claim by first proving it for some special cases. For $m=1$, the input is $\ket{x}_1\ket{0}_1$ so all we need is a single CNOT,
    \begin{flalign}
        &\xrightarrow{\mathrm{CX}(\widehat{X_0} \to Y_0)} \ket{x}_1\ket{\mathds{1}(x =0)}_1 = \ket{x}_1\ket{\text{tz}(x)}_1.&&
    \end{flalign}
    For $m=2^{i}, i\geq 0$ we prove the claim by induction on $i$. Note that we have just proved the base case of this induction, which clearly requires $O(1)$ gates and no ancilla qubits. $m=2^i$ for some $i\geq 0$ implies $\lceil\log (m+1)\rceil=i+1 =\log m +1$. 

    Assume that $\mathbf{TZ}_{m/2, \log m}$ can be implemented using $O(m/2)$ gates and no ancillas. 
    On input $\ket{x}_m\ket{0}_{\log m +1}$, apply $\mathbf{TZ}_{m/2, \log m}$ to $\ket{x_{m/2-1} \cdots x_0}_{m/2}\ket{0}_{\log m}$,
    \begin{flalign}\label{eq:appendix:TZ inductive step}
        &\xrightarrow{\mathbf{TZ}_{m/2, \log m}} \ket{x_{m-1}\cdots x_{m/2}}_{m/2}\ket{x_{m/2-1} \cdots x_{0}}_{m/2}\ket{0}_1\ket{\text{tz}(x')}_{\log m}, &&
    \end{flalign}
    where here $x' = x \bmod 2^{m/2}$. Observe that since $\text{tz}(x') \leq m/2 = 2^{\log m-1}$ it can require $\log m$ qubits to store. Since \begin{equation}\text{tz}(x) = \mathds{1}(x' =0)\text{tz}(x_{m-1} \cdots x_{m/2}) + \text{tz}(x'),\end{equation} we need to add $\text{tz}(x_{m-1} \cdots x_{m/2})$ to the $Y$ register when $x'=0$ and leave $Y$ unchanged when $x' \neq 0$.

    Suppose $x'=0$, implying $\text{tz}(x') =2^{\log m-1}$. Then we have the state
    \begin{equation}
        \ket{x_{m-1}\cdots x_{m/2}}_{m/2}\ket{0}_{m/2}\ket{0}_1\ket{1}_1\ket{0}_{\log m -1}.
    \end{equation}
    In this case, the lower half of the input register provides conditionally clean qubits \cite{khattar2025rise}. Apply the following sequence of controlled operations, to the $X$ register
    \begin{flalign}
        \xrightarrow{\mathrm{CX}(\widehat{X_{m/2}} \to  X_0)}
        & \ket{x_{m-1}\cdots x_{m/2}}_{m/2}\ket{0}_{m/2-1}\ket{\mathds{1}(x_{m/2}=0)}_1\ket{0}_1\ket{1}_1\ket{0}_{\log m -1}. &&
    \end{flalign}
    \begin{flalign}
        &\xrightarrow[j=1, \ldots, m/2-1]{\mathrm{CCX}(\widehat{X_{m/2+j}}, X_{j-1} \to  X_j)} 
        \ket{x_{m-1}\cdots x_{m/2}}_{m/2}\ket{\mathds{1}(x_{m-1} = 0 \land \cdots \land x_{m/2}=0)}_1 \cdots && \\ \nonumber & \hspace{10em} \cdots \ket{\mathds{1}(x_{m/2+1} =0 \land x_{m/2} =0)}_1\ket{\mathds{1}(x_{m/2}=0)}_1\ket{0}_1\ket{1}_1\ket{0}_{\log m -1}.&&
    \end{flalign}
    For each $j \in \Z_{m/2}$, the indicator variable on $X_{j}$ tells us if $\text{tz}(x_{m-1}\cdots x_{m/2}) \geq j+1$. We can first add $\text{tz}(x_{m-2} \cdots x_{m/2})$, which is at most $m/2-1$, to the last $\log m-1$ qubits of the $Y$ register, $Y_{\log m -2}, \ldots, Y_0$. This is possible through a sequence of controlled increments by 1, controlled on each of the qubits $X_0, \ldots, X_{m/2-2}$. This incrementer can be built with a few observations. 
    
    For an integer $j$ consider the set $S_j =\{k\geq 0: 2^{k}|(j +1)\}$. $k \in S_{j}$ if and only if the bitstrings $j$ and $j+1$ differ on their $k^{th}$ bit. Thus, to map $\ket{j}_{\log m-1} \mapsto \ket{j+1}_{\log m -1}$ for any $j \in \Z_{{m/2}-1}$ we flip $Y_k$ for each $k \in S_{j}$. Thus, the following sequence of Toffolis counts the number of trailing zeroes for $x_{m-2}\cdots x_{m/2}$,
    \begin{flalign}
        &\xrightarrow[\substack{j=0, \ldots, m/2-2, \\ k \in S_{j}}]{\mathrm{CCX}(X_{j}, Y_{\log m -1} \to Y_{k})} 
         \ket{x_{m-1}\cdots x_{m/2}}_{m/2}\ket{\mathds{1}(x_{m-1} = 0 \land \cdots \land x_{m/2}=0)}_1 \cdots &&\\ \nonumber & \cdots \ket{\mathds{1}(x_{m/2 +1} =0 \land x_{m/2} =0)}_1 
        \ket{\mathds{1}(x_{m/2}=0)}_1\ket{0}_1\ket{1}_1\ket{\text{tz}(x_{m-2}\cdots x_{m/2})}_{\log m -1}. &&
    \end{flalign}
    The importance of also controlling on $Y_{\log m -1}$ for these operations will become clear later.

    The $Y$ register is now \begin{equation}
    \text{tz}(x') + \text{tz}(x_{m-2}\cdots x_{m/2}) .
    \end{equation}
    If $X_{m/2 -1} = \ket{0}_1$, this equals $\text{tz}(x)$. If not, then $x =0$, so $\text{tz}(x) = 2^{\log m}$ and the $Y$ register contains $2^{\log m}-1$. Applying the following operations, sets $Y=\ket{\text{tz}(x)}_{\log m +1}$ for $x=0$ also.
    \begin{flalign}
        &\xrightarrow{\mathrm{CCX}(X_{m/2-1}, Y_{\log m -1} \to Y_{\log m})}
         \ket{x_{m-1}\cdots x_{m/2}}_{m/2}\ket{\mathds{1}(x_{m-1} = 0 \land \cdots \land x_{m/2}=0)}_1 \cdots&& \\ \nonumber & \ket{\mathds{1}(x_{m/2 +1} =0 \land x_{m/2} =0)}_1
        \ket{\mathds{1}(x_{m/2}=0)}_1\ket{\mathds{1}(x=0)}_1\ket{1}_1\ket{\text{tz}(x_{m-2}\cdots x_{m/2})}_{\log m -1}, &&
    \end{flalign}
    \begin{flalign}
        &\xrightarrow[\substack{j=0, \ldots, \log m -1}]{\mathrm{CX}(Y_{\log m} \to Y_j)}
        \ket{x_{m-1}\cdots x_{m/2}}_{m/2}\ket{\mathds{1}(x_{m-1} = 0 \land \cdots \land x_{m/2}=0)}_1 \cdots && \\ \nonumber & \hspace{9em} \cdots \ket{\mathds{1}(x_{m/2+1} =0 \land x_{m/2} =0)}_1\ket{\mathds{1}(x_{m/2}=0)}_1 \ket{\text{tz}(x)}_{\log m +1} &&
    \end{flalign}
    Finally, reverse the initial sequence of controlled operations that we used to compute the indicator variables. 
    \begin{flalign}
        &\xrightarrow[j=m/2-1,\ldots,1]{\mathrm{CCX}(\widehat{X_{m/2+j}}, X_{j-1}\to X_j)}
        \ket{x_{m-1}\cdots x_{m/2}}_{m/2}\ket{0}_{m/2 -1}\ket{\mathds{1}(x_{m/2} =0)}_1\ket{\text{tz}(x)}_{\log m +1}&&
    \end{flalign}
    \begin{flalign}
        &\xrightarrow{\mathrm{CX}(\widehat{X_{m/2}} \to X_0)} \ket{x_{m-1}\cdots x_{m/2}}_{m/2}\ket{0}_{m/2}\ket{\text{tz}(x)}_{\log m +1} = \ket{x}_{m}\ket{\text{tz}(x)}_{\log m +1} &&
    \end{flalign}

    Suppose instead that $x' \neq 0$ implying $\text{tz}(x) <m/2$. Returning to the input state from Equation \ref{eq:appendix:TZ inductive step}, we can clearly rewrite it as \begin{equation}
        \ket{x_{m-1}\cdots x_{m/2}}_{m/2}\ket{x_{m/2-1} \cdots x_{0}}_{m/2}\ket{0}_1\ket{0}_1\ket{\text{tz}(x')}_{\log m -1} = \ket{x}_m\ket{\text{tz}(x)}_{\log m +1}.
    \end{equation}
    Excluding the sequence of controlled operations that we used to compute and uncompute indicator variables, all of the operations in the above algorithm are either controlled on $Y_{\log m -1}$ or $Y_{\log m}$. This implies that only the initial and final operations to compute indicator variables are applied when $\text{tz}(x') <m/2$, and these cancel each other out. Thus, the state on the $Y$ register remains the same. 
    
    By induction, there exists a circuit for $\mathbf{TZ}_{m,\log m +1}$ for all $m$ satisfying $m = 2^i$ for some $i\geq 0$. The gate count of this circuit is \begin{align}G(\mathbf{TZ}_{m,\log m +1}) = G(\mathbf{TZ}_{m/2, \log m}) + O(m/2) +  O(\sum_{j=0}^{m/2-2} |S_{j}|) = O(m).\end{align} The circuit also requires no ancillas.

    Next, we prove the claim for integers $m\geq 1$ of the form $m \neq 2^{i}$ for all $i\geq 0$. In this case $m = 2^{\lfloor\log m \rfloor} + t$ for some $0<t<2^{\lfloor \log m \rfloor}$. This means $\lceil \log (m+1) \rceil = \lfloor \log m \rfloor +1$. 
    
    Let $n = 2^{\lfloor\log m \rfloor}$. On input $\ket{x}_{n+t}\ket{0}_{\log n +1}$, apply $\mathbf{TZ}_{n, \log n +1}$ to $\ket{x_{n-1} \cdots x_{0}}_{n}\ket{0}_{\log n +1}$
    \begin{flalign}
        &\xrightarrow{\mathbf{TZ}_{n, \log n +1}} \ket{x_{n+t -1} \cdots x_{n}}_{t}\ket{x_{n-1} \cdots x_0}_{n}\ket{\text{tz}(x_{n-1} \cdots x_0)}_{\log n +1}. &&
    \end{flalign}
    Then, for the remaining $t$ qubits, we apply an algorithm that is similar to the power of two case. 
    
    Suppose $\text{tz}(x_{n-1} \cdots x_0) = n$. This means our input is
    \begin{equation}
    \ket{x_{n+t-1}\cdots x_{n}}_{t}\ket{0}_{n}\ket{1}_1\ket{0}_{\log n}. 
    \end{equation}
    Apply the following operations,
    \begin{flalign}
        \xrightarrow{\mathrm{CX}(\widehat{X_{n}} \to  X_0)}
        & \ket{x_{n+t -1} \cdots x_{n}}_{t}\ket{0}_{n-1}\ket{\mathds{1}(x_{n}=0)}_1\ket{1}_1\ket{0}_{\log n}, &&
    \end{flalign}
    \begin{flalign}
        &\xrightarrow[j=1, \ldots, t-1]{\mathrm{CCX}(\widehat{X_{n+j}}, X_{j-1} \to  X_j)} 
        \ket{x_{n+t-1}\cdots x_{n}}_{t}\ket{0}_{n-t}\ket{\mathds{1}(x_{n+t-1} = 0 \land \cdots \land x_{n}=0)}_1 \cdots && \\ \nonumber & \hspace{12em} \cdots \ket{\mathds{1}(x_{n+1} =0 \land x_{n} =0)}_1\ket{\mathds{1}(x_{n}=0)}_1\ket{1}_1\ket{0}_{\log n }. &&
    \end{flalign}
    Since $\text{tz}(x_{n+t-1} \cdots x_n) \leq t <2^{\log n}$, we can apply,
    \begin{flalign}
        &\xrightarrow[\substack{j=0, \ldots, t-1, \\ k \in S_{j}}]{\mathrm{CCX}(X_{j}, Y_{\log n} \to Y_{k})} 
         \ket{x_{n+t-1}\cdots x_{n}}_{t}\ket{0}_{n-t}\ket{\mathds{1}(x_{n+t-1} = 0 \land \cdots \land x_{n}=0)}_1 \cdots &&\\ \nonumber& \hspace{8em} \cdots \ket{\mathds{1}(x_{n +1} =0 \land x_{n} =0)}_1 
        \ket{\mathds{1}(x_{n}=0)}_1\ket{1}_1\ket{\text{tz}(x_{n+t-1} \cdots x_n)}_{\log n}. &&
    \end{flalign}
    Reversing the initial controlled sequence used to compute the indicator variables, we get,
    \begin{flalign}
        &\xrightarrow[j=t-1,\ldots,1]{\mathrm{CCX}(\widehat{X_{n+j}}, X_{j-1}\to X_j)}
        \ket{x_{n+t-1}\cdots x_{n}}_{t}\ket{0}_{n -1}\ket{\mathds{1}(x_{n} =0)}_1\ket{\text{tz}(x)}_{\log n +1},&&
    \end{flalign}
    \begin{flalign}
        &\xrightarrow{\mathrm{CX}(\widehat{X_{n}} \to X_0)} \ket{x_{n+t-1}\cdots x_{n}}_{t}\ket{0}_{n}\ket{\text{tz}(x)}_{\log n +1} = \ket{x}_{n+t}\ket{\text{tz}(x)}_{\log n +1}. &&
    \end{flalign}
    
    Suppose instead $\text{tz}(x_{n-1} \cdots x_0) < n$. This implies $Y_{\log n} \neq 1$, and so the state will remain $\ket{x}_m\ket{\text{tz}(x_{n-1} \cdots x_0)}_{\log n +1}$ after the above operations. This is what we want since $\text{tz}(x) = \text{tz}(x_{n-1} \cdots x_0)$. This proves that there exists a circuit for $\mathbf{TZ}_{n+t, \log n+1}$ with gate count
    \begin{equation}
    G(\mathbf{TZ}_{n+t, \log n+1}) = G(\mathbf{TZ}_{n, \log n +1}) + O(t) + O(\sum_{j=0}^{t-1}|S_{j}|) = O(n),
    \end{equation}
    requiring no ancillas.
    
    Thus, we have proven that for all $ m \geq 1$, $\mathbf{TZ}_{m,\lceil \log (m+1) \rceil}$ can be implemented using $O(m)$ gates and no ancilla qubits.
\end{proof}

\section{Inversion mod \texorpdfstring{$\text{2}^\text{n}$}{2ⁿ}}\label{sec:appendix:inversion}
To recap, the operations $\times$ and $\div$ are defined for $x\in \Z_{2^m}^{\times}$ and $y \in \Z_{2^n}$ as follows,
\begin{align*}
    &\times(x):\ket{x}_{m}\ket{y}_n \longmapsto \ket{x}_{m}\ket{yx \bmod 2^n}_{n}, &&\div(x): \ket{x}_{m}\ket{y}_{n} \longmapsto \ket{x}_{m}\ket{y(x^{-1}) \bmod 2^n}_n.
\end{align*}
Since Lemma \ref{lem:double} also requires modular additions, we introduce the symbols $+$ and $-$ representing quantum circuits that are defined for $x \in \Z_{2^m}$ and $y \in \Z_{2^n}$ as follows,
\begin{align*}
    &+(x): \ket{x}_{m}\ket{y}_n \longmapsto \ket{x}_m\ket{y +x\bmod 2^n}_n, &&- (x): \ket{x}_{m}\ket{y}_n \longmapsto \ket{x}_m\ket{y -x \bmod 2^n}_n.
\end{align*}
As justified in the main text, for $m=O(n)$ and any $\epsilon>0$, $\times$ and $\div$ require $O(n^{1+\epsilon})$ gates and no ancilla qubits. By \cite[Sections~2.9--2.10]{gidney2017factoringdirty}, $+$ and $-$ require $O(n)$ gates and at most one dirty qubit. We drop the $\bmod \ 2^n$ notation from equations on the understanding that all operations are modular.
\InversionDouble*
\begin{proof}
    We start by proving that $y'=y+2^h((-yz)\bmod2^h)$ is the unique multiplicative inverse of $x\bmod2^{2h}$. Since $y'\equiv-2^hyz+y\pmod{2^{2h}}$, the claim follows from
\begin{align}
    x(-2^hyz + y) = xy(-2^hz +1) \equiv (1+2^hz)(-2^hz +1) \equiv 1 \pmod{2^{2h}}.
\end{align}
This motivates the first step of our algorithm. On input $\ket{z}_{m-h}\ket{y}_h$, apply
\begin{flalign}
    &\xrightarrow{\times (-y)}\ket{-yz}_{m-h}\ket{y}_{h}. &&
\end{flalign}
Thus, the least significant $2h$ qubits of the input are now $\ket{-yz \bmod 2^h}_{h}\ket{y}_h = \ket{y'}_{2h}$. Our task becomes to prepare $\ket{z'}_{m-2h}$ on the most significant $m-2h$ qubits. To do this we need to express $z'$ in terms of the current state.

Consider $v \in \Z_{2^{m-2h}}$ and $u \in \Z_{2^{h}}$ such that $-yz \bmod 2^{m-h} = 2^{h}v + u$. We claim that,
\begin{equation}\label{eq:lift}
 z'=-\frac{y'v+u^2}{y^2}\bmod2^{m-2h}.
\end{equation}
To see this, observe that $xy^2 \equiv y + 2^{h}yz  \equiv y -2^hu -2^{2h} v\pmod{2^m}$. Therefore,
\begin{equation}
\begin{split}
 y^22^{2h}z' \equiv&\; y^2(xy'-1)\\
 \equiv&\; (y-2^{h}u-2^{2h}v)(2^{h}u + y)-y^2\\
 \equiv&\; -2^{2h}u^2- 2^{2h}y'v\pmod{2^m}.
\end{split}
\end{equation}
This makes clear the algorithm we require. Rewriting the current state as $\ket{v}_{m-2h}\ket{u}_{h}\ket{y}_{h}$, we apply the following sequence of operations,
\begin{flalign}
     &\xrightarrow{\times (y')}\ket{y'v}_{m-2h}\ket u_h\ket y_h &&\\
     &\xrightarrow{+(u^2)}\ket{y'v+u^2}_{m-2h}\ket u_h\ket y_h &&\\
     &\xrightarrow{\div (y)}\ket{(y'v+u^2)/y}_{m-2h}\ket u_h\ket y_h &&\\
     &\xrightarrow{\div(-y)}\ket{-(y'v+u^2)/y^2}_{m-2h}\ket u_h\ket y_h
     =\ket{z'}_{m-2h}\ket{y'}_{2h}.
\end{flalign}
The multiplication by $y'$ uses $\ket{u}_{h}\ket{y}_{h} = \ket{y'}_{2h}$. The only operation in this algorithm which we have not yet defined is the addition by $u^2$. 

We now give an algorithm for adding $u^2$.
Let $u=2u_{\rm hi}+u_0$ for $u_0 \in \Z_{2}, u_{\rm hi} \in \Z_{2^{h-1}}$, and let $\ket{w}_{m-2h}$ be the target register for modular addition of $u^2$. 
To add $u^2$, we first add $4u_{\rm hi}^2$ using the identity
\begin{align}
    (1+2u_{\rm hi})\left(\frac{w}{1+2u_{\rm hi}}+2u_{\rm hi}\right)-2u_{\rm hi}=w+4u_{\rm hi}^2.
\end{align}
Adding $4u_0u_{\rm hi}$ and $u_0$ then gives $w+u^2$.

Our extension of Theorem~\ref{thm:main theorem} to even inputs in Appendix \ref{sec:appendix:extended main theorem} is useful here. By Corollary \ref{cor:extended quantum mod mult}, we have a circuit that acts as $\times(u)$ when $u$ is odd and as $\times(u+1)$ when $u$ is even. Both cases give the required operation $\times(1+2u_{\rm hi})$ using the existing register $\ket{u}_h$, without preparing a separate register for this odd integer. We apply the following sequence of operations:
\begin{flalign}
     &\ket w_{m-2h}\ket{u_{\rm hi}}_{h-1}\ket{u_0}_1 &&\\
    &\xrightarrow{\div(1+2u_{\rm hi})}
    \ket{w/(1+2u_{\rm hi})}_{m-2h}\ket{u_{\rm hi}}_{h-1}\ket{u_0}_1&&\\
    &\xrightarrow{+(2u_{\rm hi})}
    \ket{w/(1+2u_{\rm hi})+2u_{\rm hi}}_{m-2h}
    \ket{u_{\rm hi}}_{h-1}\ket{u_0}_1 &&\\
    &\xrightarrow{\times(1+2u_{\rm hi})}
    \ket{w+2u_{\rm hi}+4u_{\rm hi}^2}_{m-2h}
    \ket{u_{\rm hi}}_{h-1}\ket{u_0}_1 &&\\
    &\xrightarrow{-(2u_{\rm hi})}
    \ket{w+4u_{\rm hi}^2}_{m-2h}\ket{u_{\rm hi}}_{h-1}\ket{u_0}_1 &&\\
    &\xrightarrow{+(4u_0u_{\rm hi})}
    \ket{w+4u_{\rm hi}^2+4u_0u_{\rm hi}}_{m-2h}
    \ket{u_{\rm hi}}_{h-1}\ket{u_0}_1 &&\\
    &\xrightarrow{+(u_0)}
    \ket{w+4u_{\rm hi}^2+4u_0u_{\rm hi}+u_0}_{m-2h}
    \ket{u_{\rm hi}}_{h-1}\ket{u_0}_1
    =\ket{w+u^2}_{m-2h}\ket u_h.
\end{flalign}
The final equality follows from
$u^2=4u_{\rm hi}^2+4u_0u_{\rm hi}+u_0$.

We use the constructions of~\cite[Sections~2.9--2.10]{gidney2017factoringdirty}. For $\pm(2u_{\rm hi})$, we apply $\pm(u_{\rm hi})$ to all but the least significant qubit of the target $w$. For $+(4u_0u_{\rm hi})$, we apply $+(u_{\rm hi})$ to all but the least two significant target qubits, controlled on $u_0$. The operation $+(u_0)$ is a controlled increment. When required, we use $\ket{y_0}_1$ as a dirty qubit, although it is, in fact, a clean qubit, and restore it to its original state after each operation. Therefore, each addition uses $O(m-2h)$ gates and no additional qubits.

The addition by $u^2$ requires two modular multiplications and four modular additions. This brings the total for the algorithm to six modular multiplications and four modular additions. The first multiplication acts on an $(m-h)$ qubit register, and the
remaining five act on an $(m-2h)$ qubit register.
\end{proof}

\section{Main Theorem (Extended)}\label{sec:appendix:extended main theorem}

The main theorem of this work, Theorem~\ref{thm:main theorem}, involves modular multiplication with a quantum multiplier $w\in\mathbb{Z}_{2^s}^{\times}$. The restriction to odd $w$ ensures that this operation is reversible and hence unitary. For even $w$, distinct inputs can give the same output, so the operation must be modified to remain reversible.

In this section, we extend the operation defined in Theorem~\ref{thm:main theorem} to all $w\in\mathbb{Z}_{2^s}$. With some small modifications to the algorithm, an even multiplier $w$ has the same effect on the target registers as the odd multiplier $w+1$. The action for odd $w$ remains unchanged. We use this extension in the proof of Lemma \ref{lem:double} in Appendix~\ref{sec:appendix:inversion}, and it may also be of independent interest.

\begin{theorem}[Main Theorem, Extended Version]\label{thm:main theorem extended}
    Let $m>0$, $s = O(m)$ and $t=O(m)$. Let $w \in \Z_{2^s}$, $x \in \Z_{2^t}$, $y \in \Z_{2^{m}}$, $b, c \in \Z \setminus 2\Z$, and $\alpha, \beta \in \{0,1\}$. Let $w_0$ be the least significant bit of $w$ and define $f(w) = w + \mathds{1}(w_0 =0)$. For any $\epsilon >0$, there exists a quantum circuit for performing the operation \begin{equation}
    \ket{w}_{s}\ket{x}_{t}\ket{y}_{m} \longmapsto \ket{w}_{s}\ket{2^{m}bxf(w)^{\alpha}+cyf(w)^{\beta}\bmod 2^{t+m}}_{t+m}
    \end{equation}
    that uses $O(m^{1 +\epsilon})$ gates and no ancilla qubits. For the special case $s=0$, the circuit has $O(m\log m)$ depth. In this case, we adopt the convention $f(w)^{\alpha} = f(w)^{\beta} =1$.
\end{theorem}
\begin{proof}
    We use the algorithm of Theorem \ref{thm:main theorem} with some small modifications to prove the claim. For $w \in \Z_{2^s}, u \in \Z_{2^n}$ and $\phi \in \Z_{2^n}^\times$, consider the operation
    \begin{align}\label{eq:appendix:extended main theorem qft} \widetilde{\mathbf{QFT}}_{s,n}(\phi, \alpha): \ket{w}_s\ket{u}_n \longmapsto \ket{w}_s\frac{1}{\sqrt {2^{n}}} \sum_{v \in \Z_{2^n}}e^{2\pi i \phi vuf(w)^{\alpha}/2^n}\ket{\rev{v}}_n,
    \end{align}
    which is just the controlled and parametrised quantum Fourier transform when restricted to the domain $w \in \Z_{2^s}^\times, u \in \Z_{2^n}$. Using the same approach as Lemma \ref{lem:qft gate complexity}, Appendix \ref{sec:appendix:qft}, it can be shown that for $s\geq 2$,
    \begin{equation}
    \begin{split}
        \widetilde{\mathbf{QFT}}_{s,n}(\phi, 1) =& \left(\mathbf{I}_{\lceil n/2 \rceil } \otimes \widetilde{\mathbf{QFT}}_{s,\lfloor n/2 \rfloor}(\phi, 1)\right) \circ \left(\pover{0}{\lceil n/2 \rceil}{\lfloor n/2 \rfloor}{\phi/2^{n}}\right) \circ \\& \qquad \circ \left(\pover[1]{s-1}{\lceil n/2 \rceil}{\lfloor n/2 \rfloor}{2\phi/2^{n}}\right) \circ \left(\widetilde{\mathbf{QFT}}_{s,\lceil n/2 \rceil}(\phi, 1) \otimes \mathbf{I}_{\lfloor n/2 \rfloor}\right)
    \end{split}
    \end{equation}
    since $f(w) = 2\lfloor w/2 \rfloor +1$.
    This operation requires $O(n^{1+\epsilon})$ gates and no ancilla qubits, for any $\epsilon>0$.  For $s\leq 1$, $\widetilde{\mathbf{QFT}}_{s,n}(\phi, 1)$ can be implemented with $\qft{s}{n}{\phi, 0}$.
    
    Using this new operation $\widetilde{\mathbf{QFT}}_{s,n}(\phi, \alpha)$ and slightly modified integer multiplications in the phase, we apply the algorithm of Theorem \ref{thm:main theorem}.
    \begin{flalign}
    &\xrightarrow{\widetilde{\mathbf{QFT}}_{s,t}(b, \alpha) \otimes \mathbf{I}_{m}}\ket{w}_s\frac{1}{\sqrt {2^{t}}} \sum_{v \in \Z_{2^{t}}}e^{2\pi i bxvf(w)^{\alpha}/2^{t}}\ket{\rev{v}}_{t}\ket{y}_{m}, &&
    \end{flalign}
    \begin{flalign}
    &\xrightarrow{\mathbf{I}_{s} \otimes \pover{0}{t}{m}{c/2^{t+m}}} \ket{\lfloor w/2\rfloor}_{s-1}\ket{w_0}_1\frac{1}{\sqrt {2^{t}}} \sum_{v \in \Z_{2^{t}}}e^{2\pi i (bxvf(w)^{\alpha}/2^{t} + cvy/2^{t+m})}\ket{\rev{v}}_{t}\ket{y}_{m}. &&
    \end{flalign}
    If $\beta=0$ or $s\leq1$, we don't bother applying the next operation; otherwise apply,
    \begin{flalign}
    &\xrightarrow{\mathbf{I}_{1} \otimes \pover[\beta]{s-1}{t}{m}{2c/2^{t+m}}} \ket{\lfloor w/2\rfloor}_{s-1}\ket{w_0}_1\frac{1}{\sqrt {2^{t}}} \sum_{v \in \Z_{2^{t}}}e^{2\pi i (bxvf(w)^{\alpha}/2^{t} + cvyf(w)^{\beta}/2^{t+m})}\ket{\rev{v}}_{t}\ket{y}_{m}. &&
    \end{flalign}
    Then apply,
    \begin{flalign} 
    &\xrightarrow{\mathbf{I}_{t} \otimes \widetilde{\mathbf{QFT}}_{s,m}(c, \beta)} \ket{w}_{s}\frac{1}{\sqrt {2^{t+m}}} \sum_{\substack{v \in \Z_{2^{t}}\\ z \in \Z_{2^m}}}e^{2\pi i (bxvf(w)^{\alpha}/2^{t} + cvyf(w)^{\beta}/2^{t+m}+ cyzf(w)^{\beta}/2^m)}\ket{\rev{v}}_{t}\ket{\rev{z}}_{m}, &&
    \end{flalign}
    \begin{flalign}
    &\xrightarrow{\mathbf{I}_s \otimes \qft{0}{t+m}{1}^{-1}} \ket{w}_s\ket{2^{m}bxf(w)^{\alpha} + cyf(w)^{\beta}\bmod 2^{t+m}}_{t+m}.&&
    \end{flalign}
    The result follows from Lemmas \ref{lem:new P gate complexity two inputs} and \ref{lem:new P gate complexity three inputs}, and the above statements about the cost of $\widetilde{\mathbf{QFT}}_{s,n}(\phi, \alpha)$.
\end{proof}

\begin{corollary}[Modular multiplication by a quantum integer mod $2^n$, Extended Version]\label{cor:extended quantum mod mult}
    Let $z \in \Z_{2^n}$, $m=O(n)$ and $x \in \Z_{2^m}$. Let $x_0$ denote the least significant bit of $x$ and define $f(x) = x + \mathds{1}(x_0 =0)$. For any $\epsilon>0$, there exists a quantum circuit for performing the operation \begin{equation}
    \ket{x}_m\ket{z}_n \longmapsto \ket{x}_m\ket{f(x)z \bmod 2^n}_n
    \end{equation} that uses $O(n^{1 +\epsilon})$ gates and no ancilla qubits.
\end{corollary}
\begin{proof}
    Let $s=m, t = \lceil n/2\rceil, m = \lfloor n/2 \rfloor, \alpha = \beta =1,  w=x, 2^{m}x+y = z, b =c =1$ and apply Theorem \ref{thm:main theorem extended}.
\end{proof}

\end{document}